\documentclass[preprint,aps,prd,showpacs,nofootinbib,amsmath,amssymb,amsfonts,superscriptaddress]{revtex4-1}

\usepackage{amssymb}
\usepackage{bm}
\usepackage{cancel}
\usepackage{latexsym}
\usepackage{amsfonts,amssymb}
\usepackage{graphicx,epsfig}
\usepackage{psfrag}
\usepackage{amsmath,amssymb}
\usepackage{mathrsfs}
\usepackage{amsthm}
\usepackage{bm}
\usepackage{longtable}
\usepackage{lipsum}
\usepackage{array}
\usepackage{tabu}
\usepackage{multirow}
\usepackage{latexsym}
\usepackage{mathrsfs}
\usepackage{hyperref}
\usepackage{float}%To keep the figures in place
\usepackage[section]{placeins}%To keep the figures in place
\hypersetup{
    colorlinks=true,       % false: boxed links; true: colored links
    linkcolor=red,          % color of internal links
    citecolor=blue,        % color of links to bibliography
}

\newcommand{\addlabel}[1]{%
Eq.\eqrefstepcounter{equation}%
\addlabel{#1}%
\let\]\endequation }
\newcommand{\upchi}{\protect\raisebox{2.5pt}{$\chi$}}
\begin{document}
%%%
\title{Cosmological perturbations in the interacting dark sector: \\ Mapping fields and fluids}
%%%%
\author{Joseph P Johnson}
\email{josephpj@iitb.ac.in}
\affiliation{Department of Physics, Indian Institute of Technology Bombay, Mumbai 400076, India} 
\author{S. Shankaranarayanan} \email{shanki@phy.iitb.ac.in}
\affiliation{Department of Physics, Indian Institute of Technology Bombay, Mumbai 400076, India}

\begin{abstract}
There is no unique way to describe the dark energy-dark matter interaction, 
as we have little information about the nature and dynamics of the dark sector. 
Hence, in many of the phenomenological dark matter fluid interaction models in the literature, the interaction strength $Q_{\nu}$ in the dark sector is introduced by hand. Demanding that the interaction strength $Q_{\nu}$ in the dark sector must have a field theory description, we obtain a unique form of interaction strength.  We show the equivalence between the fields and fluids for the 
$f(R,\upchi)$ model where $f$ is an arbitrary, smooth function of $R$ and classical scalar field $\upchi$, which represents dark matter. Up to first order in perturbations, we show that the one-to-one mapping between the \emph{classical} field theory description and the phenomenological fluid description of interacting dark energy and dark matter exists \emph{only} for this unique form of interaction. We then classify the interacting dark energy models considered in the literature into two categories based on the field-theoretic description. We introduce a novel autonomous system {and its stability analysis} for the general interacting dark sector. We show that the dark-energy dominated epoch occurs earlier than the non-interacting systems for a specific scalar field potential and a range of coupling strengths.
\end{abstract}
\maketitle
\section{Introduction}

Dark matter dominates the galaxy mass, and dark energy forms the majority of our 
Universe's energy density ~\cite{1998-Riess.Others-Astron.J.,*1999-Perlmutter.Others-Astrophys.J.,*2007-Spergel.Others-Astrophys.J.Suppl.,*2018-Scolnic.Others-Astrophys.J.,*2018-Akrami.Others-,*2018-Aghanim.Others-}. However, we have little information about the properties
of these two components that dominate the energy content of the Universe today ~\cite{1968-Sakharov-Sov.Phys.Dokl.,*1989-Weinberg-Rev.Mod.Phys.,*2003-Padmanabhan-Phys.Rept.,*2003-Peebles.Ratra-Rev.Mod.Phys.,*2000-Sahni.Starobinsky-Int.J.Mod.Phys.,*2006-Copeland.etal-Int.J.Mod.Phys.,*2012-Bamba.etal-Astrophys.SpaceSci.,*2012-Yoo.Watanabe-Int.J.Mod.Phys.D,*2013-Tsujikawa-proc,*2016-Davis.Parkinson-proc}. 
The only information we have about the two components is that (i) Dark energy contributes
with negative pressure to the energy budget, and  (ii) Dark matter has negligible, possibly zero,
pressure ~\cite{1998-Riess.Others-Astron.J.,*1999-Perlmutter.Others-Astrophys.J.,*2007-Spergel.Others-Astrophys.J.Suppl.,*2018-Scolnic.Others-Astrophys.J.,*2018-Akrami.Others-,*2018-Aghanim.Others-}.  The above properties are based on gravitational interactions. More importantly, 
we do not know how they interact with each other and Baryons/Photons. 

In the early Universe, 
due to the tight-coupling of Baryons and Photons, the baryons participate in the acoustic
oscillations of the photons, and also cause Silk damping ~\cite{2000-Padmanabhan-TheoreticalAstrophysicsVolume,*2005-Mukhanov-PhysicalFoundationsCosmology,*2008-Weinberg-Cosmology,*2011-Gorbunov.Rubakov-IntroductionTheoryEarly}.  Near recombination, the baryons
decouple from the photons, and photons propagate freely. Solar eclipse measurements rule out dark matter interaction with
Photons.  Local gravity measurements rule out dark energy interactions with
Baryons ~\cite{2002-Sumner-LivingRev.Rel.}. However, the current observations can not constraint (or rule out) the interaction
strength between dark matter-dark energy. Interestingly, the dark matter-dark energy interaction
provides a mechanism to alleviate the coincidence problem~(see, for instance, Refs. ~\cite{2004-Farrar.Peebles-Astrophys.J.,*2014-Bolotin.etal-Int.J.Mod.Phys.D,*2016-Wang.etal-Rept.Prog.Phys.}). 
Besides this, recently, it has been shown that the dark matter-dark energy interaction can reconcile the tensions in the 
Hubble constant $H_0$ ~\cite{2017-DiValentino.etal-Phys.Rev.D,*2017-Kumar.Nunes-Phys.Rev.D,*2018-Yang.etal-Phys.Rev.Da,*2018-Yang.etal-JCAP,*2019-Pan.etal-Phys.Rev.D,*2020-DiValentino.etal-Phys.Rev.D,*2020-GomezValent.etal-}.

Naturally, there has been a surge in constructing dark energy-dark matter models~\cite{2000-Amendola-Mon.Not.Roy.Astron.Soc.,2000-Amendola-Phys.Rev.D,2000-Billyard.Coley-Phys.Rev.D,2005-Olivares.etal-Phys.Rev.D,2007-Amendola.etal-Phys.Rev.D,2008-Olivares.etal-Phys.Rev.D,2008-Boehmer.etal-Phys.Rev.D,2009-CalderaCabral.etal-Phys.Rev.D,2008-He.Wang-JCAP,2008-Pettorino.Baccigalupi-Phys.Rev.D,2008-Quartin.etal-JCAP,2010-Boehmer.etal-Phys.Rev.D,2011-Beyer.etal-Phys.Rev.D,2010-LopezHonorez.etal-Phys.Rev.D,2012-Avelino.Silva-Phys.Lett.B,2015-Pan.etal-Mon.Not.Roy.Astron.Soc.,2013-Salvatelli.etal-Phys.Rev.D,2013-Chimento.etal-Phys.Rev.D,2014-Amendola.etal-Phys.Rev.D,2016-Marra-Phys.DarkUniv.,2017-Bernardi.Landim-Eur.Phys.J.C,2017-Pan.Sharov-Mon.Not.Roy.Astron.Soc.,2018-VanDeBruck.Mifsud-Phys.Rev.D,2018-CarrilloGonzalez.Trodden-Phys.Rev.D,2019-Barros.etal-JCAP,2019-Landim-Eur.Phys.J.C}. In all these
models, phenomenologically, the interaction is proposed between the fluid terms in the dark sector. 
More specifically, individually, dark matter (DM) and dark energy (DE) 
do not satisfy the conservation equations, however, the combined sector satisfies the energy conservation equation ~\cite{2004-Farrar.Peebles-Astrophys.J.,*2014-Bolotin.etal-Int.J.Mod.Phys.D,*2016-Wang.etal-Rept.Prog.Phys.}, i. e.,
\begin{equation}
\nabla^{\mu} {T}_{\mu \nu }^{({\rm DE, DM})} 
= Q_{\nu}^{({\rm DE,DM})} 
\label{eq:IntDEDM}
\end{equation}
such that 
\begin{equation}
Q_{\nu}^{({\rm DE})}  + Q_{\nu}^{({\rm DM})}  = 0
\label{def:Q}
\end{equation}
where $Q$ determines the interaction strength between
dark matter and dark energy. Since the gravitational effects on dark matter and dark energy are opposite, even a small interaction can impact the cosmological evolution~\cite{2014-Bolotin.etal-Int.J.Mod.Phys.D}. Since we have little information about the dark sector, in many of these models, the interaction strength $Q_{\nu}$ in the dark sector is put in by hand. 

However, it is unclear whether these broad classes of phenomenological models can be obtained from a field theory action. More specifically, can the above interaction strength $Q_{\nu}$ in the dark sector be derived systematically from a field theory action. Attempts have been made in the literature to obtain the interaction strength from the field-theoretic action~\cite{2018-CarrilloGonzalez.Trodden-Phys.Rev.D}. The correspondence between the fluid description and field-theoretic description is established \emph{only} for the background cosmology and not for the perturbations. The analysis of cosmological perturbations is essential to provide a complete understanding 
of these models and, more importantly, to determine if the perturbations are stable in the presence of interaction $Q_{\nu}$. 

In this work, we show the equivalence up to first order in the perturbations of $f(R,\upchi)$ model where $f$ is an arbitrary, smooth function of $R$ and the classical scalar field $\upchi$ which represents dark matter. More specifically, under conformal transformations, we show that $f(R, \upchi)$ is equivalent to a model with two coupled scalar fields. The coupling between the classical scalar fields, which gives rise to the dark energy - dark matter interaction \eqref{eq:IntDEDM} can be represented by the evolution equations of the dark energy (represented by a scalar field) and dark matter (represented by a fluid). We show that the interaction between the dark sectors can be rewritten in terms of the trace of the energy-momentum tensor of the dark matter fluid, and a coupling function depending on the dark energy field. We then look at several interacting dark sector models that are proposed in the literature and identify the compatible models with the field theory action proposed here. 

We define a set of dimensionless variables and construct an autonomous system that completely describes the dark energy - dark matter interaction and background evolution. We analyze the fixed points of the system and show that the system has a stable attractor solution, corresponding to the late-time accelerated expansion of the Universe. To our knowledge, this is the first time such an approach is used to study a general class of interacting dark sector models. We consider a specific dark energy-dark matter interaction model and study the background evolution. We show that for a range of (both positive and negative) coupling strengths, the dark-energy dominated epoch occurs earlier with an interacting dark sector than in the non-interacting dark sector. 

In this work we use the natural units where $c=1, \kappa^2 = 8 \pi G$, and the metric signature $(-,+,+,+)$. Greek alphabets denote the 4-dimensional space-time coordinates and Latin alphabets denote the 3-dimensional spatial coordinates. Overbarred quantities (like $\overline{\rho}(t),  \overline{P}(t)$) are evaluated for the FRW background and a \textit{dot} represents the derivative with respect to cosmic time $t$. Unless otherwise specified, subscript ``$, \phi$" denotes derivative with respect to $\phi$, 
subscript ``$, \upchi$" denotes derivative with respect to $\upchi$,  and 
subscript $m$ denotes dark matter.
 
\section{Dark sector interaction from a field theory action}
\label{sec:2}

In the field theory description of the interacting dark energy - dark matter models, the coupling between the dark sector components is represented by a coupling term, which is an arbitrary function of the dark energy scalar field.
It can be shown that modified gravity models such as $f(\tilde{R},\tilde{\upchi})$ gravity can lead to such models~~\cite{2019-Johnson.etal-Gen.Rel.Grav.}.
Consider the following action in Jordan frame:
\begin{equation}
\label{eq:fRaction}
S_{J}=\int d^{4}x\sqrt{-\tilde{g}}\left[\frac{1}{2 \kappa^2} f(\tilde{R},\tilde{\upchi})- \frac{1}{2}   \tilde{g}^{\mu \nu} \tilde{\nabla}_{\mu} \tilde{\upchi} \tilde{\nabla}_{\nu} \tilde{\upchi} -V(\tilde{\upchi})\right] 
\end{equation}
where $f(\tilde{R},\tilde{\upchi})$ is an arbitrary, smooth function of Ricci scalar, and scalar field $\tilde{\upchi}$, and $V(\upchi)$ is the self-interaction potential of the scalar field $\tilde{\upchi}$. Under the conformal transformation: 
\begin{equation}
\label{eq:conftrans}
{g}_{\mu \nu}=\Omega^{2} \tilde{g}_{\mu \nu},
\quad \mbox{where} \quad
\Omega^{2}= F(\tilde{R},\tilde{\upchi}) \equiv \frac{\partial f(\tilde{R}, \tilde{\upchi})}{\partial \tilde{R}} 
\end{equation}
and a field redefinition, the action in the Einstein frame takes the following form 
\begin{equation}
 \label{eq:Scde}
S = \int d^4x \sqrt{-g}\left(\dfrac{1}{2 \kappa^2}R-\dfrac{1}{2}g^{\mu \nu}\nabla_{\mu}\phi \nabla_{\nu}\phi - U(\phi)-\dfrac{1}{2} e^{2\alpha(\phi)}g^{\mu \nu}\nabla_{\mu}\upchi \nabla_{\nu}\upchi -e^{4 \alpha(\phi)} V(\upchi) \right).
\end{equation}
where 
\[
 U = \frac{F\tilde{R}-f}{2\kappa^2 F^{2}} \, .
\]
This action has also been considered in the context of a multi-field inflationary scenario. (See, for instance, Refs~\cite{2001-Starobinsky.etal-Nucl.Phys.B,*2003-DiMarco.etal-Phys.Rev.D}.) Recently, the same action is also considered in Ref. \cite{2018-CarrilloGonzalez.Trodden-Phys.Rev.D}. However, to our knowledge, we have not seen an explicit calculation that shows the derivation of the above action in the Einstein frame. Appendix~\ref{app:A} contains the details of the transformations in the field space to derive the above action. 

From the above action \eqref{eq:Scde}, the field equations for $\upchi$ and $\phi$, respectively, are:
\begin{eqnarray}
\label{eq:eomchi}
-\nabla^{\mu}\nabla_{\mu}\upchi  - 2 \alpha_{, \phi}(\phi) \nabla_{\mu}\phi \nabla^{\mu}\upchi +e^{2 \alpha(\phi)}V_{, \chi}(\upchi)&=& 0 \\
\label{eq:eomphi}
-\nabla^{\mu}\nabla_{\mu}\phi + 4e^{4\alpha}\alpha_{, \phi}(\phi)V(\upchi)+ e^{2\alpha}\alpha_{,\phi}(\phi)\nabla^{\mu}\upchi \nabla_{\mu}\upchi+
U_{, \phi}(\phi) &=& 0
\end{eqnarray}
where the notations such as $V_{,\chi}, U_{,\phi}$ denote $\partial V/\partial 
\upchi, \partial U/\partial \phi$.  The variation of action \eqref{eq:Scde} with respect to the metric $g_{\mu\nu}$ gives the Einstein's equation 
\begin{equation}
\label{eq:fieldeq}
G_{\mu \nu} = \kappa^2 T_{\mu\nu} \, ,
\end{equation}
where the stress-tensor is given by:
\begin{equation}
\label{eq:stresstensortotal}
T_{\mu\nu} = 
\nabla_{\mu}\phi \nabla_{\nu}\phi - \dfrac{1}{2}g_{\mu \nu}\nabla^{\sigma}\phi \nabla_{\sigma}\phi 
- g_{\mu \nu}U(\phi)  + e^{2\alpha}\nabla_{\mu}\upchi \nabla_{\nu}\upchi
- \dfrac{1}{2}e^{2\alpha} g_{\mu \nu} \nabla^{\sigma}\upchi \nabla_{\sigma}\upchi - e^{4\alpha}g_{\mu \nu}V(\upchi) \, .
\end{equation}
In the field theoretic description, the two field equations \eqref{eq:eomchi}, \eqref{eq:eomphi} and the Einstein's equation \eqref{eq:fieldeq} completely describe the system.  

Since the dark matter and the dark energy constitute up to $95 \%$ of the energy content of the Universe today, it is good a approximation to assume that the total energy momentum tensor of the Universe is given by  \eqref{eq:stresstensortotal}. Demanding the local conservation of the energy-momentum tensor leads to:
\begin{equation}
\label{eq:etot}
\nabla^{\mu} T_{\mu \nu} = \nabla^{\mu} T_{\mu \nu}^{(\phi)} +
 \nabla^{\mu} T_{\mu \nu}^{(\upchi)} = 0 \, .
\end{equation}
where $T_{\mu \nu}^{(\phi)}$ and $T_{\mu \nu}^{(\upchi)}$ refer to the stress-tensor corresponding to scalar fields $\phi$ and $\upchi$, respectively.  
Due to the interaction between the two fields $\phi$ and $\upchi$, there is no unique way to write the stress-tensor corresponding to the scalar fields, and the conservation of the energy momentum tensor of the individual components is violated.  Following \eqref{eq:IntDEDM}, \eqref{eq:eomchi} and \eqref{eq:eomphi}, the interaction between the two scalar fields can be described as:
\begin{equation}
\label{eq:dedmint}
-\nabla^{\mu} {T}_{\mu \nu }^{(\phi)}= Q_{\nu}^{\rm (F)}  = 
\nabla^{\mu} {T}_{\mu \nu }^{(\upchi)}
\end{equation}
where 
\begin{eqnarray}
\label{eq:emtensor}
T^{(\upchi)}_{\mu \nu} &=&  e^{2\alpha(\phi)}\left(\nabla_{\mu}\upchi \nabla_{\nu}\upchi - \dfrac{1}{2}g_{\mu \nu} \nabla^{\sigma} \upchi \nabla_{\sigma} \upchi - e^{2 \alpha(\phi)}g_{\mu \nu}V(\upchi)\right) \\
T^{(\phi)}_{\mu \nu} &=&\nabla_{\mu}\phi \nabla_{\nu}\phi-\dfrac{1}{2}g_{\mu \nu}\nabla^{\sigma}\phi \nabla_{\sigma}\phi-g_{\mu \nu}U(\phi) \\
\label{eq:interaction}
Q_{\nu}^{\rm (F)}  &=& \nabla^{\mu} T_{\mu\nu}^{(\upchi)} =  -e^{2\alpha(\phi)} \alpha_{,\phi}(\phi) \nabla_{\nu} \phi \left[ \nabla^{\sigma} \upchi \nabla_{\sigma} \upchi + 4 e^{2\alpha(\phi)} V(\upchi) \right]
\end{eqnarray}
It is important to note that starting from \eqref{eq:fRaction}, we can obtain 
interaction strength $Q^{\rm (F)} $ in terms of $\phi$ and  $V(\upchi)$. We can equally rewrite $Q^{\rm (F)} $ in-terms of $U(\phi)$. While this field theory description may be considered a fundamental description of the system, 
the fluid description turns out to be more useful to analyze the cosmological observations. In that regard, the most common description of the interacting dark sector is in terms of dark matter fluid.

\subsection{Fluid description of the interacting dark sector}
\label{sec:2a}

In the fluid description, it is often convenient to consider the dark matter to be fluid. For this purpose, we replace the dark matter scalar field and related quantities by the corresponding energy density $\rho_m$, pressure $p_m$  of the dark matter fluid~\cite{2018-CarrilloGonzalez.Trodden-Phys.Rev.D}:
\begin{equation}
\label{eq:dmrhop}
p_{m}  = -\dfrac{1}{2}e^{2 \alpha}\left[g^{\mu \nu} \nabla_{\mu} \upchi \nabla_{\nu} \upchi + e^{2\alpha}V(\upchi) \right], \quad
\rho_{m}  = -\dfrac{1}{2}e^{2 \alpha}\left[g^{\mu \nu} \nabla_{\mu} \upchi \nabla_{\nu} \upchi - e^{2\alpha}V(\upchi) \right].
\end{equation} 
The four velocity $u_{\mu}$ of the dark matter fluid is given by
\begin{equation}
\label{eq:dm4v}
u_{\mu} = -\left[-g^{\alpha \beta} \nabla_{\alpha}\upchi \nabla_{\beta}\upchi \right]^{-\frac{1}{2}} \nabla_{\mu} \upchi
\end{equation}
In this description, the Einstein's equation can be rewritten in terms of dark energy scalar field and dark matter fluid:
\begin{equation}
\label{eq:flfeq}
 G_{\mu \nu} = \kappa^2 \left[\nabla_{\mu}\phi \nabla_{\nu}\phi - \dfrac{1}{2}g_{\mu \nu} \nabla^{\sigma}\phi \nabla_{\sigma}\phi -g_{\mu \nu} V(\phi)+ p_m g_{\mu \nu} + (\rho_m + p_m) u_{\mu} u_{\nu} \right] \, ,
\end{equation}
where the energy-momentum tensor for the dark matter fluid is given by
\begin{equation}
T^{(m)\mu}_{\nu}=p_m g_{\mu \nu} + (\rho_m + p_m) u_{\mu} u_{\nu} \, ,
\end{equation}
and the interaction term can be rewritten as
\begin{equation}
\label{eq:interaction02}
Q_{\nu}^{\rm (F)} = \nabla_{\mu} T^{(m)\mu}_{\nu} =  -e^{2\alpha(\phi)} \alpha_{,\phi}(\phi) \nabla_{\nu} \phi \left[ \nabla^{\sigma} \upchi \nabla_{\sigma} \upchi + 4 e^{2\alpha(\phi)} V(\upchi) \right] =  -\alpha_{,\phi}(\phi) \nabla_{\nu}\phi (\rho_m  - 3 p_m) 
\end{equation}
Identifying $T^{(m)}=T^{(m) \mu}_{\mu} = -(\rho_m - 3 p_m ) $, we get
\begin{equation}
\label{eq:traceinter}
Q_{\nu}^{\rm (F)}  = T^{(m)} \nabla_{\nu}\alpha(\phi)
\end{equation}
Thus, we see that in the fluid description of interacting dark matter, the interaction term is proportional to the trace of the energy-momentum tensor of the dark matter and the coupling $\alpha$. It is important to note that starting from the Jordan frame action \eqref{eq:fRaction}, the form of the interaction term $Q_{\nu}^{\rm (F)}$ is \emph{uniquely} written in terms of dark energy scalar field and dark matter fluid. 

This has to be contrasted with the dark matter interaction fluid models in the literature  Refs. ~\cite{2000-Amendola-Mon.Not.Roy.Astron.Soc.,2000-Amendola-Phys.Rev.D,2000-Billyard.Coley-Phys.Rev.D,2005-Olivares.etal-Phys.Rev.D,2007-Amendola.etal-Phys.Rev.D,2008-Olivares.etal-Phys.Rev.D,2008-Boehmer.etal-Phys.Rev.D,2009-CalderaCabral.etal-Phys.Rev.D,2008-He.Wang-JCAP,2008-Pettorino.Baccigalupi-Phys.Rev.D,2008-Quartin.etal-JCAP,2010-Boehmer.etal-Phys.Rev.D,2011-Beyer.etal-Phys.Rev.D,2010-LopezHonorez.etal-Phys.Rev.D,2012-Avelino.Silva-Phys.Lett.B,2015-Pan.etal-Mon.Not.Roy.Astron.Soc.,2013-Salvatelli.etal-Phys.Rev.D,2014-Amendola.etal-Phys.Rev.D,2016-Marra-Phys.DarkUniv.,2017-Bernardi.Landim-Eur.Phys.J.C,2017-Pan.Sharov-Mon.Not.Roy.Astron.Soc.,2018-VanDeBruck.Mifsud-Phys.Rev.D,2018-CarrilloGonzalez.Trodden-Phys.Rev.D,2019-Barros.etal-JCAP,2019-Landim-Eur.Phys.J.C})  where ${Q}_{\nu}$ can take any form. In the next section we show that a one-to-one correspondence between the fields and the fluids is only true if the interaction term is given by $Q_{\nu}^{\rm (F)}$ in Eq. \eqref{eq:traceinter}. 

\section{Cosmological evolution with dark energy - dark matter interaction}
\label{sec:3}

To study the cosmological evolution with interacting dark sector, we consider the spatially flat FRW metric with first order scalar perturbations in synchronous gauge \footnote{For the evolution equations in Newtonian gauge, see Appendix ~\ref{App2}}~\cite{2008-Weinberg-Cosmology}:
\begin{equation}
\label{eq:syncmetric}
g_{00}=-1,\quad g_{0i}=0, \quad g_{ij} = a^2 \left[(1+A)\delta_{ij}+ \dfrac{\partial^2 B}{\partial x^i \partial x^j}\right].
\end{equation}
where $a\equiv a(t)$ is the scale factor with Hubble parameter given by $H = \dot{a}/a$ and $A\equiv A(t,x,y,z)$ and $B\equiv B(t,x,y,z)$ are scalar perturbations. At the linear order, the scalar, vector and tensor perturbations decouple, and can be treated separately. Since the scalar perturbations couple to the energy density $(\delta \rho)$ and pressure $(\delta P)$ leading to the growing inhomogeneities, we only consider scalar perturbations. 

The scalar fields $\phi$ and $\upchi$, dark matter fluid energy density ($\rho_{m}$), dark matter fluid pressure ($p_m$) and the interaction strength ($Q_{\nu}$) can be split into background and perturbed parts as:
\begin{equation}
\label{eq:rhompmdef}
\phi = \overline{\phi}+\delta\phi ,\quad \upchi = \overline{\upchi} + \delta \upchi, \quad \rho_{m}
=\overline{\rho}_m+\delta \rho_m, \quad p_m = \overline{p}_m+\delta p_m, 
\quad Q_{\nu} = \overline{Q}_{\nu} + \delta Q_{\nu}
\end{equation} 
Usually in the literature, in the fluid description, the dark matter is assumed to be pressureless dust, i. e.  $\overline{p}_m=\delta p_m=0$. In this work, we 
\emph{do not} {make this assumption} for the dark matter fluid, i. e., 
$\overline{p}_m \neq 0 $ and $\delta p_m \neq 0$. Although, all our calculations are valid in the special case of pressureless dust.

Components of dark matter fluid four velocity can be written as:
\begin{equation}
\label{eq:dmfourv}
u_{\mu} = \overline{u}_{\mu}+\delta u_{\mu}, \quad \overline{u}_0 = -1, \quad \delta u_0 = 0 , \quad \overline{u}_i=0 , \quad \delta u_i = \dfrac{\partial \delta u^s}{\partial x^i} , \quad \delta u^s = -\dfrac{\delta \upchi}{\dot{\overline{\upchi}}}
\end{equation}
In the following subsections, we present the evolution equations for the background and the first-order perturbations. 

\subsection{Correspondence between fields and fluids in the FRW background}
\label{sec:3a}

In the fluid description, the Friedmann equations for the interacting dark sector are given by~\cite{2016-Wang.etal-Rept.Prog.Phys.}:
\begin{eqnarray}
\left(\dfrac{\dot{a}}{a} \right)^2 &=& \dfrac{\kappa^2}{3} \left(\overline{\rho}_m + \dfrac{\dot{\overline{\phi}}^2}{2} + U(\overline{\phi}) \right) \nonumber \\
2 \dfrac{\ddot{a}}{a} + \left(\dfrac{\dot{a}}{a}\right)^2 &=& -\dfrac{\kappa^2}{3} \left(\overline{p}_m + \dfrac{\dot{\overline{\phi}}^2}{2}-U(\overline{\phi}) \right).
\end{eqnarray}
From Eq. \eqref{eq:IntDEDM}, the conservation equations for the dark energy field and dark matter fluid in the FRW background are given by:
\begin{eqnarray}
 \label{eq:dmflbg}
 \ddot{\overline{\phi}}\dot{\overline{\phi}}+3H\dot{\overline{\phi}}^2+U_{,\phi}(\overline{\phi})\dot{\overline{\phi}}  &=& \overline{Q}  \nonumber \\
\dot{\overline{\rho}}_m + 3 H (\overline{\rho}_m + \overline{p}_m) &=& -\overline{Q} \, .
\end{eqnarray}
{In the phenomenological description of the dark matter fluid interaction}, there is no unique form of $\overline{Q}$. Several authors have considered many different forms of $\overline{Q}$ in the literature (See, for instance, Refs. ~\cite{2000-Amendola-Mon.Not.Roy.Astron.Soc.,2000-Amendola-Phys.Rev.D,2000-Billyard.Coley-Phys.Rev.D,2005-Olivares.etal-Phys.Rev.D,2007-Amendola.etal-Phys.Rev.D,2008-Olivares.etal-Phys.Rev.D,2008-Boehmer.etal-Phys.Rev.D,2009-CalderaCabral.etal-Phys.Rev.D,2008-He.Wang-JCAP,2008-Pettorino.Baccigalupi-Phys.Rev.D,2008-Quartin.etal-JCAP,2010-Boehmer.etal-Phys.Rev.D,2011-Beyer.etal-Phys.Rev.D,2010-LopezHonorez.etal-Phys.Rev.D,2012-Avelino.Silva-Phys.Lett.B,2015-Pan.etal-Mon.Not.Roy.Astron.Soc.,2013-Salvatelli.etal-Phys.Rev.D,2014-Amendola.etal-Phys.Rev.D,2016-Marra-Phys.DarkUniv.,2017-Bernardi.Landim-Eur.Phys.J.C,2017-Pan.Sharov-Mon.Not.Roy.Astron.Soc.,2018-VanDeBruck.Mifsud-Phys.Rev.D,2018-CarrilloGonzalez.Trodden-Phys.Rev.D,2019-Barros.etal-JCAP,2019-Landim-Eur.Phys.J.C}). However, as discussed in Sec. \eqref{sec:2a}, starting from the Jordan frame action \eqref{eq:fRaction}, the interaction term $Q_{\nu}^{\rm (F)} $ in Eq. \eqref{eq:traceinter}  is uniquely written in terms of dark energy scalar field and dark matter fluid. In this case, 
the background interaction term is given by
\begin{equation}
\overline{Q}^{\rm (F)}  = -\alpha_{,\phi}(\overline{\phi}) \dot{\overline{\phi}}(\overline{\rho}_m - 3\overline{p}_m) \, .
\end{equation} 
We now show that the above equations are consistent with the field theory description \emph{only} for this form of interaction term $\overline{Q}^{\rm (F)}$.  Using the definition of $p_{m}$ and $\rho_{m}$ in Eq.~\eqref{eq:dmrhop}, the evolution equations for the scalar field $\phi$ and $\upchi$ are given by:
\begin{eqnarray}
\label{eq:eomchibg}
\ddot{\overline{\upchi}} + 3H\dot{\overline{\upchi}} + e^{2\alpha}V_{,\chi}(\overline{\upchi}) + 2 \alpha_{,\phi}(\overline{\phi})\dot{\overline{\phi}}\dot{\overline{\upchi}}&=& 
0 \nonumber \\
\label{eq:eomphiflbg}
\ddot{\overline{\phi}} + 3H\dot{\overline{\phi}} + U_{,\phi}(\overline{\phi}) + 4 e^{4\alpha} \alpha_{,\phi}(\overline{\phi})V(\overline{\upchi})-e^{2\alpha}\alpha_{,\phi}(\overline{\phi})\dot{\overline{\upchi}}^2 &=& 0 \, .
\end{eqnarray}
The background interaction term in the field theory picture also can be obtained by the direct substitution of the variables:
\begin{equation}
\overline{Q}^{\rm (F)} = \alpha_{,\phi}(\overline{\phi})\dot{\overline{\phi}}e^{2\alpha(\overline{\phi})}\left[\dot{\overline{\upchi}}^2 - 4 e^{2\alpha} V(\overline{\upchi}) \right] 
\end{equation}
Similarly, the Friedmann equations in the field theory description can be obtained by substituting $\overline{\rho}_m$ and $\overline{p}_m$ with corresponding field theory variables. From the above analysis, it is clear that there is a one-to-one correspondence between the fluids and fields \emph{only} for interaction term $\overline{Q}^{\rm (F)}$. For any other form of the interaction term, the correspondence may not exist. In Sec. \eqref{sec:4}, we
classify various models used in the literature based on this 
correspondence.

\subsection{Correspondence between fields and fluids in first order perturbations}
\label{sec:3b}

In the fluid description, the first order scalar perturbations, in synchronous gauge, satisfy the following equations~\cite{2008-Weinberg-Cosmology}:
\begin{eqnarray}
\dot{A} &=& \kappa^2 \left[(\overline{p}_m+\overline{\rho}_m)\delta u^s - \dot{\overline{\phi}}\delta \phi \right]  \\
\ddot{B}+3H\dot{B} - \dfrac{A}{a^2} &=& 0 \\
\dfrac{3}{2}\ddot{A}+\nabla^2 \left[\dfrac{1}{2} \ddot{B} + H \dot{B} \right] + 3 H \dot{A} &=& 
\dfrac{\kappa^2}{2} \left[-\delta \rho_m - 3 \delta p_m - 4 \dot{\overline{\phi}}\dot{\delta \phi} + 2 U_{,\phi}(\overline{\phi}) \delta \phi \right]  \\
- \dfrac{1}{2}\ddot{A}+\dfrac{1}{2 a^2} \nabla^2 A -3 H \dot{A} - \dfrac{1}{2}H \nabla^2 \dot{B} &=& \dfrac{\kappa^2}{2} \left[ -\delta \rho_m + \delta p_m - 2 U_{,\phi}(\overline{\phi}) \delta \phi\right]
\end{eqnarray}
From Eq. \eqref{eq:IntDEDM}, the conservation equations for the dark energy field and dark matter fluid in the first order perturbations are given by:
\begin{eqnarray}
\dot{\delta \rho_m} + 3 H (\delta p_m + \delta \rho_m)+(\overline{p}_m+\overline{\rho}_m)\left[\dfrac{\nabla^2 \delta u^s}{a^2} + \dfrac{3}{2}\dot{A} + \dfrac{\nabla^2 \dot{B}}{2}\right] &=& -\delta Q \\
\dot{\overline{\phi}}\left(\ddot{\delta \phi}-\dfrac{\nabla^2 \delta \phi}{a^2} + U_{,\phi \phi}(\overline{\phi}) \delta \phi \right) + \dot{\delta \phi} \left( \ddot{\overline{\phi}}+6 H \dot{\overline{\phi}} + U_{,\phi}(\overline{\phi}) \right) + \dfrac{\dot{\overline{\phi}}^2}{2} \left( \nabla^2 \dot{B} + 3 \dot{A} \right) &=& \delta Q  \, .
\end{eqnarray}
The above equations are generic equations for the coupled dark matter fluid
and dark energy field with arbitrary interaction term $\delta {Q}$. As mentioned earlier, there is no unique form of $\delta {Q}$ in the phenomenological description of the dark matter fluid interaction. Several authors have considered many different forms of $\delta {Q}$ in the literature (See, for instance, Refs. ~\cite{2000-Amendola-Mon.Not.Roy.Astron.Soc.,2000-Amendola-Phys.Rev.D,2000-Billyard.Coley-Phys.Rev.D,2005-Olivares.etal-Phys.Rev.D,2007-Amendola.etal-Phys.Rev.D,2008-Olivares.etal-Phys.Rev.D,2008-Boehmer.etal-Phys.Rev.D,2009-CalderaCabral.etal-Phys.Rev.D,2008-He.Wang-JCAP,2008-Pettorino.Baccigalupi-Phys.Rev.D,2008-Quartin.etal-JCAP,2010-Boehmer.etal-Phys.Rev.D,2011-Beyer.etal-Phys.Rev.D,2010-LopezHonorez.etal-Phys.Rev.D,2012-Avelino.Silva-Phys.Lett.B,2015-Pan.etal-Mon.Not.Roy.Astron.Soc.,2013-Chimento.etal-Phys.Rev.D,2013-Chimento.etal-Phys.Rev.D,2013-Salvatelli.etal-Phys.Rev.D,2014-Amendola.etal-Phys.Rev.D,2016-Marra-Phys.DarkUniv.,2017-Bernardi.Landim-Eur.Phys.J.C,2017-Pan.Sharov-Mon.Not.Roy.Astron.Soc.,2018-VanDeBruck.Mifsud-Phys.Rev.D,2018-CarrilloGonzalez.Trodden-Phys.Rev.D,2019-Barros.etal-JCAP,2019-Landim-Eur.Phys.J.C}). However, as discussed in Sec. \eqref{sec:2a}, starting from the Jordan frame action \eqref{eq:fRaction}, the interaction term $Q_{\nu}^{\rm (F)} $ in Eq. \eqref{eq:traceinter}  is uniquely written in terms of dark energy scalar field and dark matter fluid. In this case, the perturbed interaction term is given by
\begin{equation}
 \delta Q^{\rm (F)} =-(\delta \rho_m - 3 \delta p_m)\alpha_{,\phi}(\overline{\phi})\dot{\overline{\phi}}-(\overline{\rho}_m-3\overline{p}_m)\left[\alpha_{,\phi \phi}(\overline{\phi}) \dot{\overline{\phi}} \delta \phi + \alpha_{,\phi}(\overline{\phi}) \dot{\delta \phi}\right] 
\end{equation}
Like in the previous subsection, we now show that the above equations are consistent with the field theory description \emph{only} for this form of interaction ${Q}^{\rm (F)}$.  Substituting $\rho_{m}, p_{m},\delta \rho_{m}$, and $\delta p_{m}$ from Eq.~\eqref{eq:dmrhop}, the perturbed equations of motion for $\phi$ and $\upchi$, respectively, are:
\begin{eqnarray} 
\ddot{\delta \upchi}-\dfrac{\nabla^2 \delta \upchi}{a^2} + e^{2\alpha}V_{,\chi \chi}(\overline{\upchi}) \delta \upchi + \dfrac{\dot{\overline{\upchi}}}{2} \left( \nabla^2 \dot{B} + 3 \dot{A} \right) + 3 H \dot{\delta \upchi} + 2 \alpha_{,\phi}(\overline{\phi}) \left(\dot{\overline{\phi}} \dot{\delta \upchi}
 + \dot{\overline{\upchi}} \dot{\delta \phi}\right) & & \nonumber \\
%%%%
  + 2 \delta \phi \left[ \dot{\overline{\phi}}\dot{\overline{\upchi}} \alpha_{,\phi \phi}(\overline{\phi}) + e^{2 \alpha} \alpha_{,\phi}(\overline{\phi}) V_{,\chi}(\overline{\upchi}) \right] &=&  0 \\
%%%%%%%%%%%%%
\ddot{\delta \phi}-\dfrac{\nabla^2 \delta \phi}{a^2} + U_{,\phi \phi}(\overline{\phi}) \delta \phi + \dfrac{\dot{\overline{\phi}}}{2} \left( \nabla^2 \dot{B} + 3 \dot{A} \right) + 2e^{2\alpha} \alpha_{,\phi}(\overline{\phi})\left[2 e^{2\alpha} V_{,\chi}(\overline{\upchi}) \delta \upchi - \dot{\overline{\upchi}} \dot{\delta \upchi} \right]  & &  \nonumber \\
%%%%
 + 2 e^{2\alpha} \alpha_{,\phi}(\overline{\phi})^2 \delta \phi \left[8 e^{2\alpha} V(\overline{\upchi}) - \dot{\overline{\upchi}}^2 \right]+e^{2\alpha}\alpha_{,\phi \phi}(\overline{\phi}) \delta \phi \left[4 e^{2\alpha}V(\overline{\upchi}) - \dot{\overline{\upchi}}^2 \right] &=& 0
\end{eqnarray}
The above perturbed field equations are identical to the equations obtained from Eqs. (\ref{eq:eomchi}, \ref{eq:eomphi}), respectively. The perturbed interaction term in the field theory picture also can be obtained by the direct substitution of the variables:
\begin{eqnarray} \nonumber
\delta Q^{\rm (F)} = 2 e^{2\alpha}\alpha_{,\phi}(\overline{\phi}) \dot{\overline{\phi}}\left[\dot{\overline{\upchi}}\dot{\delta \upchi}-2 e^{2 \alpha} V_{,\chi}(\overline{\upchi}) \delta \upchi \right] + e^{2\alpha} \alpha_{,\phi \phi}(\overline{\phi}) \dot{\overline{\phi}} \delta \phi \left[ \dot{\overline{\upchi}}^2 - 4 V(\overline{\upchi}) \right] 
\\
+ 2 e^{2 \alpha} \alpha_{,\phi}(\overline{\phi})^2 \dot{\overline{\phi}} \delta \phi \left[ \dot{\overline{\upchi}}^2 - 8 e^{2\alpha} V(\overline{\upchi})\right] + e^{2\alpha} \alpha_{,\phi}(\overline{\phi}) \dot{\delta \phi}\left[\dot{\overline{\upchi}}^2 - 4 e^{2 \alpha} V(\overline{\upchi}) \right]
\end{eqnarray}
We would like to stress the following points regarding the above results: First, there is no unique form of $\delta {Q}$ in the phenomenological description of the dark matter fluid interaction. However, demanding a one-to-one correspondence between the field and fluid picture leads to a unique interaction term $Q_{\nu}^{\rm (F)}$. Second, we see that apart from the convenience of relating the variable to cosmological observables, evolution equations in the fluid description are simpler than those in the fluid theory description, making it easier for the numerical analysis of the model. Third, while the form of the interaction term is unique, it still contains unknown functions like $\alpha(\phi)$, $\upchi$ and $V(\upchi)$. In the next section, we now use this correspondence to clarify the phenomenological dark matter fluid interaction models in the literature~\cite{2000-Amendola-Mon.Not.Roy.Astron.Soc.,2000-Amendola-Phys.Rev.D,2000-Billyard.Coley-Phys.Rev.D,2005-Olivares.etal-Phys.Rev.D,2007-Amendola.etal-Phys.Rev.D,2008-Olivares.etal-Phys.Rev.D,2008-Boehmer.etal-Phys.Rev.D,2009-CalderaCabral.etal-Phys.Rev.D,2008-He.Wang-JCAP,2008-Pettorino.Baccigalupi-Phys.Rev.D,2008-Quartin.etal-JCAP,2010-Boehmer.etal-Phys.Rev.D,2011-Beyer.etal-Phys.Rev.D,2010-LopezHonorez.etal-Phys.Rev.D,2012-Avelino.Silva-Phys.Lett.B,2015-Pan.etal-Mon.Not.Roy.Astron.Soc.,2013-Chimento.etal-Phys.Rev.D,2013-Salvatelli.etal-Phys.Rev.D,2014-Amendola.etal-Phys.Rev.D,2016-Marra-Phys.DarkUniv.,2017-Bernardi.Landim-Eur.Phys.J.C,2017-Pan.Sharov-Mon.Not.Roy.Astron.Soc.,2018-VanDeBruck.Mifsud-Phys.Rev.D,2018-CarrilloGonzalez.Trodden-Phys.Rev.D,2019-Barros.etal-JCAP}.

\section{Interacting dark energy models in the literature}
\label{sec:4}
Since we have little information about the nature and dynamics of the dark sector, there is no unique way of describing the interaction between dark energy and dark matter. Till now, the interaction strength $Q_{\nu}$ is described by phenomenological models, with model parameters constrained by cosmological observations ~\cite{2004-Farrar.Peebles-Astrophys.J.,*2014-Bolotin.etal-Int.J.Mod.Phys.D,*2016-Wang.etal-Rept.Prog.Phys.}.  
In many of the models, the interaction strength $Q_{\nu}$ in the dark sector is constructed using the energy densities of the dark energy and dark matter, and other dynamic quantities appearing in the model. However, it is not clear whether the models can be written from a field-theoretic action. 

In this work, starting from the Jordan frame action \eqref{eq:fRaction}, we showed that the interaction term $Q_{\nu}^{\rm (F)}$ is unique. We showed that this interaction provides a one-to-one mapping between the field and fluid description of the dark matter sector. Armed with this, in this section, we classify the interacting dark energy models considered in the literature into two categories based on the field-theoretic description.  The table below identifies the models that can (or can not) be described by the field theory approach considered in this work. The list is not exhaustive but gives a good representation of the various models discussed in the literature.

     \begin{longtable}{|r|c|c|}      
      \hline
\textbf{Interacting DE-DM} & \textbf{DE-DM Interaction} & \textbf{Is }\\
      \textbf{model}   & $\nabla^{\mu} {T}_{\mu \nu }^{({\rm DE, DM})} 
= Q_{\nu}^{({\rm DE,DM})} $ & 
\boldmath{$Q_{\nu} \propto Q_{\nu}^{\rm (F)}$}? \\
      \hline
      Amendola - 1999 \citep{2000-Amendola-Mon.Not.Roy.Astron.Soc.} & $\dot{\rho}_{m}+3 H \rho_{m}=-C \rho_{m} \dot{\phi} $& Yes\\
      \hline
      Amendola - 1999 \citep{2000-Amendola-Phys.Rev.D} & $\dot{\rho}_{m}+3 H \rho_{m}=-C \rho_{m} \dot{\phi} $& Yes\\
      \hline
      Billyard \& Coley -1999 \citep{2000-Billyard.Coley-Phys.Rev.D} & $ \dot{\phi}(\ddot{\phi}+3 H \dot{\phi}+k V)=\frac{(4-3 \gamma)}{2 \sqrt{\omega+\frac{3}{2}}} \dot{\phi} \mu $ & Yes \\
      \hline
      Olivares.etal - 2005 \citep{2005-Olivares.etal-Phys.Rev.D} & $\frac{d \rho_{c}}{d t}+3 H \rho_{c}=3 H c^{2}\left(\rho_{c}+\rho_{x}\right)$ & No \\
      \hline
      Amendola.etal - 2006 \citep{2007-Amendola.etal-Phys.Rev.D} & $\dot{\rho}_{D M}+3 H \rho_{D M}-\delta(a) H \rho_{D M}=0$ & No \\
      \hline
      Olivares.etal - 2007 \citep{2008-Olivares.etal-Phys.Rev.D} & $\dot{\rho}_{c}+3 H \rho_{c}=3 H c^{2}\left(\rho_{x}+\rho_{c}\right)$ & No \\
      \hline
      \multirow{2}{*}{Boehmer.etal - 2008 \citep{2008-Boehmer.etal-Phys.Rev.D}} & $\dot{\rho}_{c}+3 H \rho_{c}=-\sqrt{2 / 3} \kappa \beta \rho_{c} \dot{\varphi}$ & Yes \\ & $\dot{\rho}_{c}+3 H \rho_{c}=-\alpha H \rho_{c}$ &  No \\
\hline
      \multirow{2}{*}{Caldera-Cabral.etal - 2008 \citep{2009-CalderaCabral.etal-Phys.Rev.D}} & $\dot{\rho}_{c}=-3 H \rho_{c}+3 H\left(\alpha_{x} \rho_{x}+\alpha_{c} \rho_{c}\right)$ & No \\ & $\dot{\rho}_{c}=-3 H \rho_{c}+3\left(\Gamma_{x} \rho_{x}+\Gamma_{c} \rho_{c}\right)$ & No \\
\hline
       \multirow{2}{*}{He \& Wang - 2008 \citep{2008-He.Wang-JCAP}} & $\dot{\rho}_{D M}+3 H \rho_{D M}-\delta H \rho_{D M}=0$ & No \\ & $\dot{\rho}_{D M}+3 H \rho_{D M}-\delta H \left(\rho_{D M}+\rho_{DE}\right)=0$ & No  \\
\hline
        Pettorino \& Baccigalupi - 2008 \citep{2008-Pettorino.Baccigalupi-Phys.Rev.D} & $\phi^{\prime \prime}+2 \mathcal{H} \phi^{\prime}+a^{2} U_{, \phi}=a^{2} C_{c} \rho_{c}$ & Yes \\
\hline
        Quartin.etal - 2008 \citep{2008-Quartin.etal-JCAP} & $\frac{\mathrm{d} \rho_{c}}{\mathrm{d} N}+3 \rho_{c}=3\lambda_{x} \rho_{x}+\lambda_{c} \rho_{c}$ & No \\
\hline
        \multirow{3}{*}{Boehmer.etal - 2009 \citep{2010-Boehmer.etal-Phys.Rev.D}} & $\dot{\rho}_{c}=-3 H \rho_{c}-\frac{\alpha}{M_{0}} \rho_{\varphi}^{2}$ & No \\ & $\dot{\rho}_{c}=-3 H \rho_{c}-\frac{\beta}{M_{0}} \rho_{c}^{2}$ & No  \\ & $\dot{\rho}_{c}=-3 H \rho_{c}-\frac{\gamma}{M_{0}} \rho_{\varphi} \rho_{c}$ & No  \\
\hline
       Beyer.etal - 2010 \citep{2011-Beyer.etal-Phys.Rev.D} & $\ddot{\varphi}+3 H \dot{\varphi}-\alpha M^{3} e^{-\alpha \varphi / M}=\frac{\beta}{M} \rho_{\chi}$ & Yes  \\
\hline
       Lopez Honorez.etal - 2010 \citep{2010-LopezHonorez.etal-Phys.Rev.D} & $\dot{\rho}_{d m}+3 H \rho_{d m}=\beta(\phi) \rho_{d m} \dot{\phi}$ & Yes  \\
\hline
      Avelino \& Silva - 2012 \citep{2012-Avelino.Silva-Phys.Lett.B} & $\dot{\rho}_{m}+3 H \rho_{m}=\alpha H a^{\beta}\rho_w$ & No  \\
\hline
      Pan.etal - 2012 \citep{2015-Pan.etal-Mon.Not.Roy.Astron.Soc.} & $\dot{\rho}_{m}+3 H \rho_{m}=3 \lambda_{m} H \rho_{m}+3 \lambda_{d} H \rho_{d}$ & No  \\
\hline
      Salvatelli.etal - 2013 \citep{2013-Salvatelli.etal-Phys.Rev.D} & $\dot{\rho}_{d m}+3 \mathcal{H} \rho_{d m}=\xi \mathcal{H} \rho_{d e}$ & No  \\
\hline
      Chimento.etal - 2013 \citep{2013-Chimento.etal-Phys.Rev.D} & $\rho_{\mathrm{m}}^{\prime}+\gamma_{\mathrm{m}} \rho_{\mathrm{m}}= -\alpha \rho^{\prime} \rho$ & No \\
\hline
      Amendola.etal - 2014 \citep{2014-Amendola.etal-Phys.Rev.D} & $\dot{\rho}_{\alpha}+3 H \rho_{\alpha}=-\kappa \sum_{i} C_{i \alpha} \dot{\phi}_{i} \rho_{\alpha}$ & Yes  \\
\hline
      Marra - 2015 \citep{2016-Marra-Phys.DarkUniv.} & $\dot{\rho}_{m}+3 H \rho_{m}= \nu \delta^n_m \rho_m \dot{\phi}/M_{Pl}$ & No  \\
\hline
      \multirow{2}{*}{Bernardi \& Landim - 2016 \citep{2017-Bernardi.Landim-Eur.Phys.J.C}} & $\dot{\rho}_{m}+3 H \rho_{m}=Q \left(\rho_{\phi}+\rho_m \right) \dot{\phi}$ & No \\ & $\dot{\rho}_{m}+3 H \rho_{m}=Q \rho_{\phi} \dot{\phi}$ & No  \\
\hline
      Pan \& Sharov - 2016 \citep{2017-Pan.Sharov-Mon.Not.Roy.Astron.Soc.} & $\dot{\rho}_{d m}+3 \mathcal{H} \rho_{d m}=3 \lambda_{m} H \rho_{dm}+3 \lambda_{d} H \rho_{d}$ & No  \\
\hline
      \multirow{2}{*}{Bruck \& Mifsud - 2017 \citep{2018-VanDeBruck.Mifsud-Phys.Rev.D}}\footnote{Violates causality condition  ($D(\phi)>0$) for the disformal transformations~\citep{1993-Bekenstein-Phys.Rev.D}} & $\nabla^{\mu} T_{\mu \nu}^{D M}=Q \nabla_{\nu} \phi$ & Yes \\ & $Q=\frac{C_{, \phi}}{2 C} T_{D M}+\frac{D_{, \phi}}{2 C} T_{D M}^{\mu \nu} \nabla_{\mu} \phi \nabla_{\nu} \phi-\nabla_{\mu}\left[\frac{D}{C} T_{D M}^{\mu \nu} \nabla_{\nu} \phi\right]$ & if $D=0$  \\
\hline
      Gonzalez \& Trodden - 2018 \citep{2018-CarrilloGonzalez.Trodden-Phys.Rev.D} & $\dot{\rho}_{\chi}+3 H \rho_{\chi}=\alpha^{\prime} \dot{\phi} \rho_{\chi}$ & Yes  \\
\hline
      Barros.etal - 2018 \citep{2019-Barros.etal-JCAP} & $\dot{\rho}_{c}+3 H \rho_{c}=-\kappa \beta \dot{\phi} \rho_{.}$ & Yes  \\
\hline
      Landim - 2019 \cite{2019-Landim-Eur.Phys.J.C} & $\ddot{\phi}+3 H \dot{\phi}+V^{\prime}(\phi)=-Q \rho_{m}$ & Yes \\
\hline
%    \end{tabular}
\end{longtable}
%  \end{center}
%\end{table}
\section{Detailed analysis of background evolution}

This section considers the class of interacting models with a one-to-one mapping between the fields and fluids and shows that one can exactly solve the evolution equation of the dark matter fluid. We do a detailed analysis of the background evolution of the dark sector in the fluid picture. 

The background interaction term \eqref{eq:traceinter} in the fluid picture can be rewritten as
\begin{equation}
\label{eq:qredef}
\overline{Q}^{\rm (F)}  = -\alpha_{,\phi}(\overline{\phi}) \dot{\overline{\phi}}(\overline{\rho}_m - 3\overline{p}_m) = -\dot{\alpha}(\overline{\phi}) (\overline{\rho}_m - 3\overline{p}_m)
\end{equation}
Assuming a matter fluid with a constant equation of state $\omega_{m} = \overline{p}_m / \overline{\rho}_m$, we can rewrite the continuity equation for matter fluid as
\begin{equation}
\dot{\overline{\rho}}_m + 3 H \overline{\rho}_m(1 + \omega_m) =\dot{\alpha}(\overline{\phi}) \overline{\rho}_m(1 - 3\omega_m)
\end{equation}
Solving  we get:
\begin{equation}
\label{eq:rhoevo}
\overline{\rho}_m = \overline{\rho}_{m_0} a ^{-3(1+\omega_m)}e^{[\alpha(\overline{\phi})-\alpha_0](1-3\omega_m)},
\end{equation}
where $\overline{\rho}_{m_0}$ and $\overline{\phi}_0$ are the current values of $\overline{\rho}_m$ and $\overline{\phi}$ and $\alpha_0 = \alpha(\overline{\phi}_0)$. The evolution of $\phi$ is determined by the equation of motion of $\phi$.
As expected, setting $\alpha(\phi)=0$, results in the evolution of $\rho_m$ in the $\Lambda$CDM model. 

\subsection{Autonomous system of interacting dark energy-dark matter model}
\label{subsec:autosys}
To study the cosmological evolution in the interacting dark-sector, we write the equations in dimensionless variables and describe them as an autonomous system of equations~\cite{1998-Copeland.etal-Phys.Rev.D,2006-Copeland.etal-Int.J.Mod.Phys.}. To our knowledge, this is the first time such an approach is used to study a general class of interacting dark sector models. 

To study and analyse a general class of dark sector interaction models, we define the following dimensionless variables:
\begin{equation}
x = \sqrt{\dfrac{C_1}{6}}\dfrac{\dot{\phi}}{H M_{Pl}}, \quad y = \sqrt{\dfrac{C_1}{3}}\dfrac{\sqrt{U}}{H M_{Pl}}
\end{equation}
\begin{equation}
\lambda = - \dfrac{M_{Pl}}{\sqrt{C_1}}\dfrac{U_{,\phi}}{U}, \quad \Gamma = \dfrac{U U_{,\phi \phi}}{U_{,\phi}^2}
\end{equation}
\begin{equation}
\alpha = \alpha(\phi), \quad \beta =  -\dfrac{M_{Pl}}{\sqrt{C_1}}\dfrac{\alpha_{,\phi}}{\alpha}, \quad \gamma = \dfrac{\alpha \alpha_{,\phi \phi}}{\alpha_{,\phi}^2}
\end{equation}
where \textit{dot} represents derivative w.r.t. time, and $C_1$ is a constant. ($ C_1$ is defined in Appendix~\ref{App:3}.) Here $\alpha,\beta$ and $\gamma$ describes a general interaction function and its properties.
These variables can be used to study a large class of interacting dark energy - dark matter models. 
The following equations give the autonomous system of the interacting dark-sector model:
\begin{eqnarray}
x' + \dfrac{3}{2}x \left(1 - x^2 + y^2 - \dfrac{\Omega_r}{3} \right) - \sqrt{\dfrac{3}{2}}\left( \lambda y^2 + \dfrac{q}{x} \right) &=& 0
\\
y' + \dfrac{3}{2} y \left( \sqrt{\dfrac{2}{3}}\lambda x - x^2 + y^2 - \dfrac{\Omega_r}{3} - 1 \right) &=& 0
\\
\Omega_m' + \Omega_m \left(3 y^2 -3 x^2  -\Omega_r \right) + \sqrt{6} q  &=& 0
\\
\Omega_r' + \Omega_r \left(1 - 3 x^2 + 3 y^2 - \Omega_r \right) &=& 0
\\
\lambda' + \sqrt{6}\lambda^2 x \left( \Gamma - 1 \right) &=& 0
\\
\beta' + \sqrt{6} \beta^2 x (\gamma-1) &=& 0
\\
\alpha' + \sqrt{6} \alpha \beta x &=& 0
\end{eqnarray}
and the energy constraint is given by
\begin{equation}
x^2 + y^2 + \Omega_m + \Omega_r - 1 = 0
\end{equation}
Here \textit{prime} denotes derivative with respect to the number of e-foldings $N \equiv \ln(a)$.
For the pressureless matter fluid, the scaled interaction term ($q$) is defined as
\begin{equation}
\label{eq:qQdef}
q \equiv \alpha \, \beta \, x \, \Omega_m = - \dfrac{ \alpha_{,\phi}(\overline{\phi}) \dot{\overline{\phi}} \overline{\rho}_m}{3 \sqrt{6}H^3 M_{Pl}^2} = \dfrac{\overline{Q}}{3 \sqrt{6}H^3 M_{Pl}^2}.
\end{equation}
Note that various cosmological parameters can be expressed in terms of these variable as
\begin{eqnarray}
 \Omega_{\phi} = x^2 + y^2, & \quad &  \omega_{\phi} = \dfrac{x^2 - y^2}{x^2 + y^2} \\
%%%
\label{def:epsilon}
\rho_i = 3 H^2 M_{Pl}^2 \Omega_i, & \quad &  \epsilon \equiv -\dfrac{\dot{H}}{H^2} = \dfrac{3}{2} \left( x^2 - y^2 + \dfrac{\Omega_r}{3} + 1 \right)
\end{eqnarray}

\subsection{Stability analysis of the autonomous system}
\label{sec:stability}
To get further insight into the background evolution of the Universe with the interacting dark sector, we look at the fixed points of the autonomous system introduced in Sec. \eqref{subsec:autosys}. We consider two cases: 
\begin{itemize}
\item {\bf Case (i)}: Models with constant $\lambda$ (exponential scalar field potential) and linear interaction function (constant $\alpha \beta$). 
\item {\bf Case (ii)}:  Models with general scalar field potential ($\lambda \neq constant$) and a general coupling function ($\alpha \beta \neq constant$). To our knowledge, a general stability analysis for this case has not been done before.
\end{itemize}
Table~\eqref{tab:fix_point} contains the list of fixed points of the autonomous system for case (i). 
\LTcapwidth=\textwidth
\renewcommand{\arraystretch}{1.5}
 \begin{longtable}{|c|c|c|c|c|c|}      
     \hline
Fixed point & \textbf{$x^*$} & \textbf{$y^*$} & \textbf{$\Omega_{r}^*$ }&$\Omega_{\phi}^*$& $\epsilon^*$\\
   \hline
1a & -1 & 0 & 0 & 1 & 3 \\
\hline
1b & 0 & 0 & 1 & 0 & 2 \\
\hline
1c & 1 & 0 & 0 & 1 & 3 \\
\hline
1d & $\dfrac{1}{\sqrt{6}\alpha \beta}$ & 0 & $1-\dfrac{1}{2\alpha^2 \beta^2}$ & $\dfrac{1}{6 \alpha^2 \beta^2}$ & 2 \\
\hline
1e & $\sqrt{\dfrac{2}{3}}\alpha \beta$ & 0 & 0 & $\dfrac{2}{3}\alpha^2 \beta^2$ & $\dfrac{3}{2}+\alpha^2\beta^2$ \\
\hline
1f & $-\sqrt{\dfrac{3}{2}}\dfrac{1}{\alpha \beta - \lambda}$ & $ \sqrt{\dfrac{3}{2}+\alpha^2 \beta^2 - \alpha \beta \lambda}\dfrac{1}{\alpha \beta - \lambda} $ & 0 & $ \dfrac{\alpha^2 \beta^2 -\alpha \beta \lambda + 3}{(\lambda - \alpha \beta)^2}$ & $-\dfrac{3 \lambda}{2 (\alpha \beta - \lambda)}$ \\
\hline 
1g & $\sqrt{\dfrac{2}{3}}\dfrac{2}{\lambda}$ & $\dfrac{2}{\sqrt{3} \lambda}$ & $1 - \dfrac{4}{\lambda^2}$ & $\dfrac{4}{\lambda^2}$ & 2 \\
\hline
1h & $\dfrac{\lambda}{\sqrt{6}}$ & $\sqrt{1-\dfrac{\lambda^2}{6}}$ & 0 & 1 & $\dfrac{\lambda^2}{2}$ \\
\hline
\caption{Fixed points of the autonomous system with a given $\lambda$ and linear coupling function.}
\label{tab:fix_point}
 \end{longtable} 
As mentioned above, the fixed points are for the case of exponential scalar field potential and a linear interaction function. The fixed points can be considered as instantaneous fixed points for other potentials and interaction functions. For constant $\lambda$ and $\alpha \beta$, it has been shown that a sequence of radiation dominated era, matter-dominated era, and an accelerated attractor (like, 1b $\rightarrow$ 1e $\rightarrow$ 1h) is cosmologically viable for a range of parameter values ~\cite{2006-Copeland.etal-Int.J.Mod.Phys.,2015-Amendola.Tsujikawa-DarkEnergyTheory}.

Table~\eqref{tab:fixpoint_gen} contains the list of fixed points of the autonomous system for case (ii) --- 
which is for system with evolving $\lambda$ and a general coupling function $\alpha(\phi)$(for which $\lambda$ and $\alpha \beta$ are not necessarily constants). 
\renewcommand{\arraystretch}{1.5}
\LTcapwidth=\textwidth
 \begin{longtable}{|c|c|c|c|c|c|c|c|c|c|}     
      \hline
Fixed point & \textbf{$ \quad x^* \quad$} & \textbf{$ \quad y^* \quad $} & \textbf{$ \quad \Omega_{r}^* \quad $ }&$\quad \Omega_{\phi}^* \quad $& $\quad \Omega_m^* \quad $&$\quad \lambda^* \quad $ &$\quad \alpha^* \quad $ & $\quad \beta^* \quad $& $\quad \epsilon^* \quad $\\
   \hline
2a & 0 & 0 & 1 & 0 & 0 & - & - & - & 2 \\
\hline
2b & 0 & 0 & 0 & 0 & 1 & - & 0 & - & $\dfrac{3}{2}$ \\
\hline
2c & 0 & 0 & 0 & 0 & 1 & - & - & 0 & $\dfrac{3}{2}$ \\
\hline
2d & 0 & 1 & 0 & 1 & 0 & 0 & - & - & 0 \\
\hline
2e & -1 & 0 & 0 & 1 & 0 & 0 & - & 0 & 3 \\
\hline
2f & 0 & 0 & 0 & 0 & 1 & 0 & - & 0 & $\dfrac{3}{2}$ \\
\hline
2g & 0 & 0 & 1 & 0 & 0 & 0 & - & 0 & 2 \\
\hline
2h & 0 & 1 & 0 & 1 & 0 & 0 & - & 0 & 0 \\
\hline
2i & 1 & 0 & 0 & 1 & 0 & 0 & - & 0 & 3 \\
\hline
\caption{Fixed points of the autonomous system with varying $\lambda$ and a general coupling function $\alpha$.}
 \label{tab:fixpoint_gen} 
 \end{longtable} 
In some cases, the fixed point conditions are satisfied for any physically realizable values of the parameters. These are represented as empty cells in the table. We can classify the fixed points in Table \eqref{tab:fixpoint_gen} into radiation dominated, matter-dominated, and late-time accelerated phase fixed points.
 \begin{enumerate}
 \item Radiation dominated phase: We have two radiation dominated fixed points, `2a' and `2g'. Looking at the eigenvalues of the Jacobian matrix of the system, we see that both of them are saddle points.
 \item matter-dominated phase: Matter dominated era can be realized by `2b', `2c', and `2f', and all of them are saddle points.
 \item Accelerated phase: Looking at the values of $\epsilon^*$, we see that dark energy dominated accelerated expansion can be realized by `2d' and `2h'. Both of these fixed points are attractors.
 \end{enumerate}
 From the above analysis, it is clear that the interacting dark sector model can lead to a radiation dominated era followed by a matter dominated era, followed by an accelerated phase (e.g., 2a $\rightarrow$ 2c $\rightarrow$ 2d). The attractor behavior of the accelerated fixed point ensures that the late Universe stays in the accelerated phase, leading to the de-Sitter Universe, which is indicated by $\epsilon^* = 0 \Rightarrow H^* = constant$.

\subsection{Dark energy-dark matter interaction: A specific example}

In the above subsections, we have shown that the interacting dark-sector model can be expressed 
as autonomous system. However, the analysis for an arbitrary model is not possible. Here, we consider 
quintessence dark energy model~\cite{2013-Pavlov.etal-Phys.Rev.D} with a linear interaction function:
\begin{equation}
C_1 = \dfrac{M_{Pl}^2}{2}, \quad U(\phi) = \dfrac{8\pi M_{Pl}^2\kappa}{2}\dfrac{1}{\phi}, \quad \alpha(\phi) = \dfrac{C}{\sqrt{2}}\phi
\end{equation}
where $\kappa$ and coupling strength $C$ are constants. In the rest of this section, we consider the background evolution for two different scenarios:  $C \geq 0$ and $C \leq 0$. For both the scenarios, 
we solve the above set of equations numerically in the redshift range $1500 < z < 0$, and the evolution of
various cosmological parameters are plotted w.r.t the number of e-foldings ($N$).

For the background evolution, we choose the following initial conditions and parameter values:
\begin{equation*}
\label{eq:inicon1}
x_i= 1.5 \times 10^{-5}, \quad y_i = 2.5 \times 10^{-5}, \quad \Omega_{r_i}=0.4, \quad \Omega_{m_i} = 1-x_i^2-y_i^2 - \Omega_{r_i}
\end{equation*}
\begin{equation*}
\label{eq:inicon2}
\lambda_i = 0.6, \quad \alpha_i = C/\lambda_i, \quad \beta_i = -\lambda_i, \quad \Gamma = 2, \quad \gamma = 0.
\end{equation*}
The initial values are chosen so that the evolution is consistent with the observed values of the cosmological parameters. It is important to note that a range of initial conditions will lead to the accelerated expansion of the Universe with a dark energy dominated phase. These specific initial conditions are chosen as representative values for the background evolution. The values of $\Gamma$ and $\gamma$ are fixed by the choice of the dark energy scalar field potential $U$ and the coupling function $\alpha$, respectively.

\subsubsection{Scenario I : $C\geq0$}

\begin{figure}[!htb]
\begin{minipage}[b]{.5\textwidth}
\includegraphics[scale=0.45]{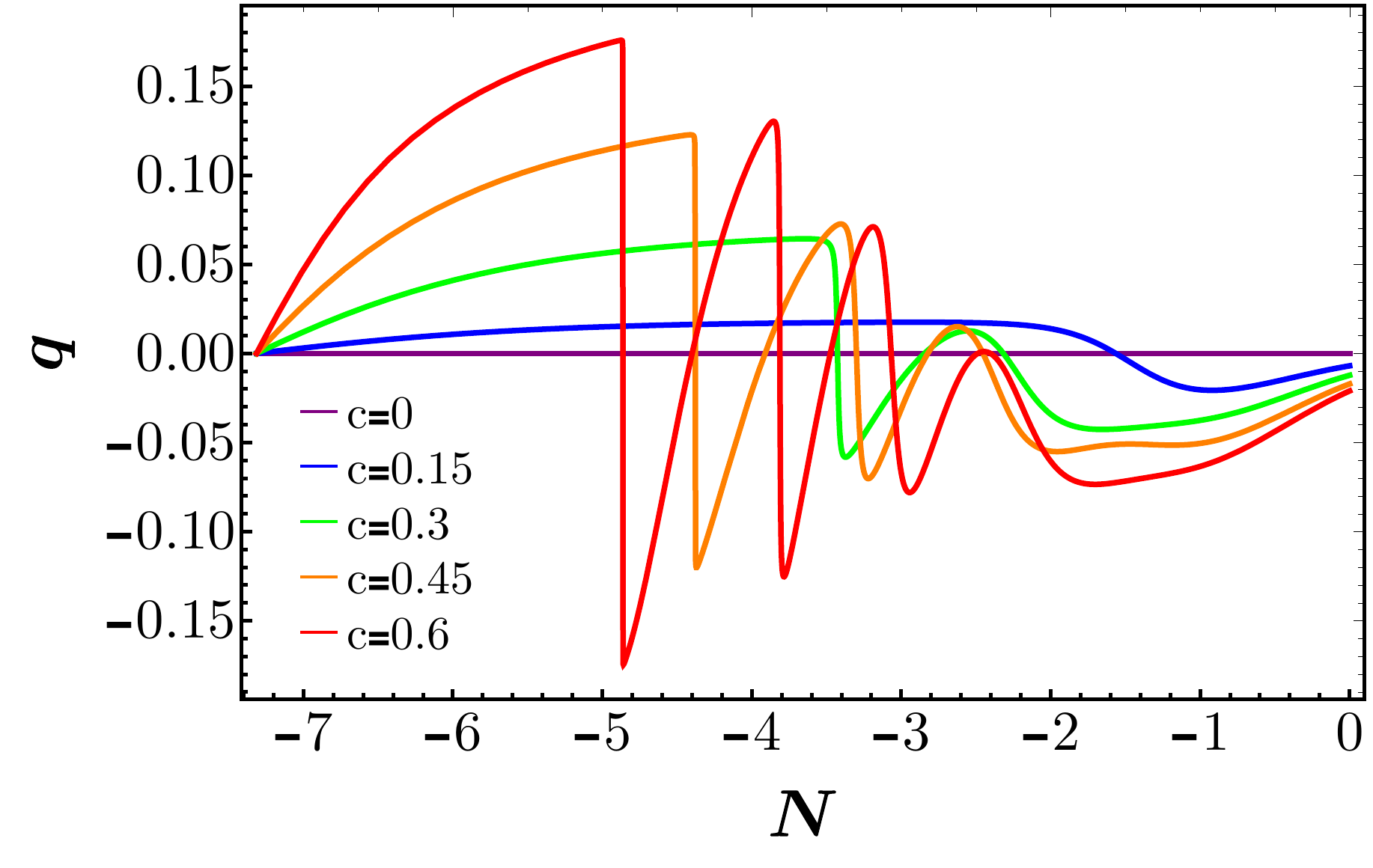}
%\caption{$k=5H_0$}\label{label-k5}
\end{minipage}\hfill
\begin{minipage}[b]{.5\textwidth}
\includegraphics[scale=0.45]{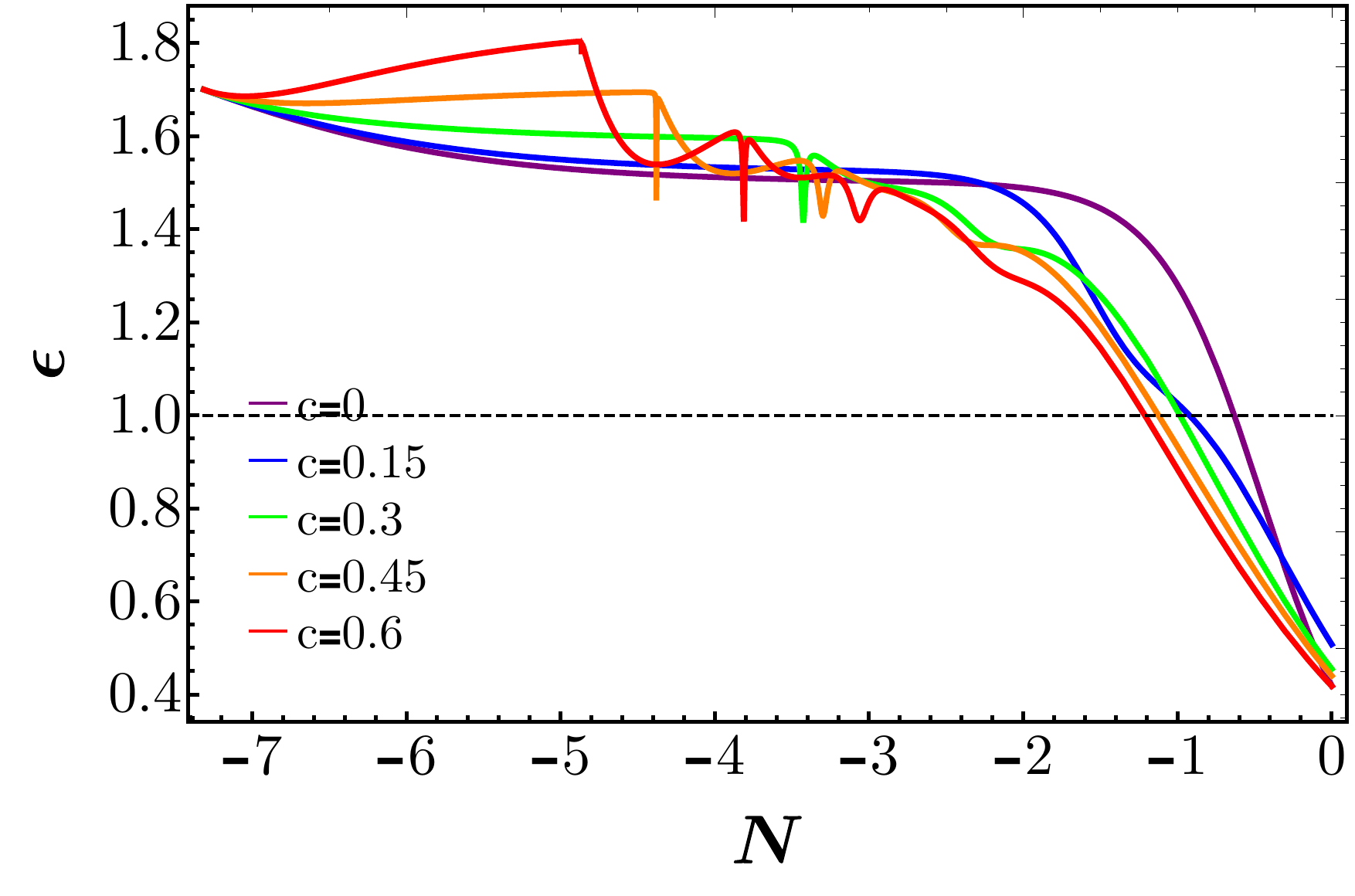}
\end{minipage}
\caption{Left panel: Evolution of interaction term $q \equiv \alpha \beta x \Omega_{m}$ as a function of $N$; Right panel: slow-roll parameter $\epsilon$ as function of $N$}
\label{fig:qeps}
\end{figure}

Fig. \eqref{fig:qeps} contains the plots of the evolution of the scaled interaction term $q$ (defined in Eq. \ref{eq:qQdef}) and slow-roll parameter $\epsilon$ (defined in Eq. \ref{def:epsilon}).
Here we see that the interaction term takes both positive and negative values during the evolution, and the strength of interaction decreases in the late Universe. All the cases result in the late-time accelerated expansion ($\epsilon < 1$). The interacting dark-sector model leads to an early dark energy dominated phase compared to the non-interacting dark-sector.

\begin{figure}[!htb]
\begin{minipage}[b]{.5\textwidth}
\includegraphics[scale=0.32]{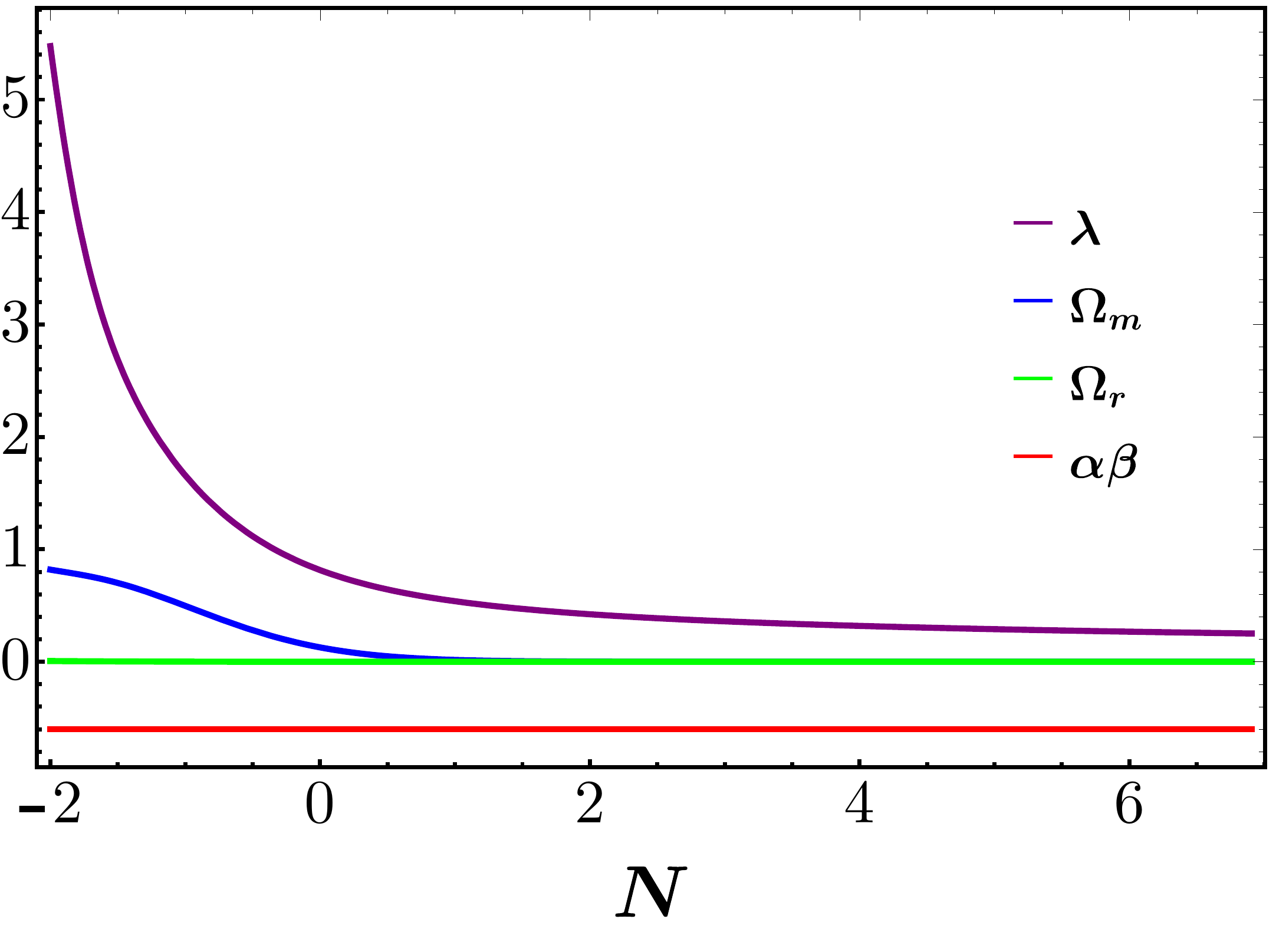}
%\caption{$k=5H_0$}\label{label-k5}
\end{minipage}\hfill
\begin{minipage}[b]{.5\textwidth}
\includegraphics[scale=0.32]{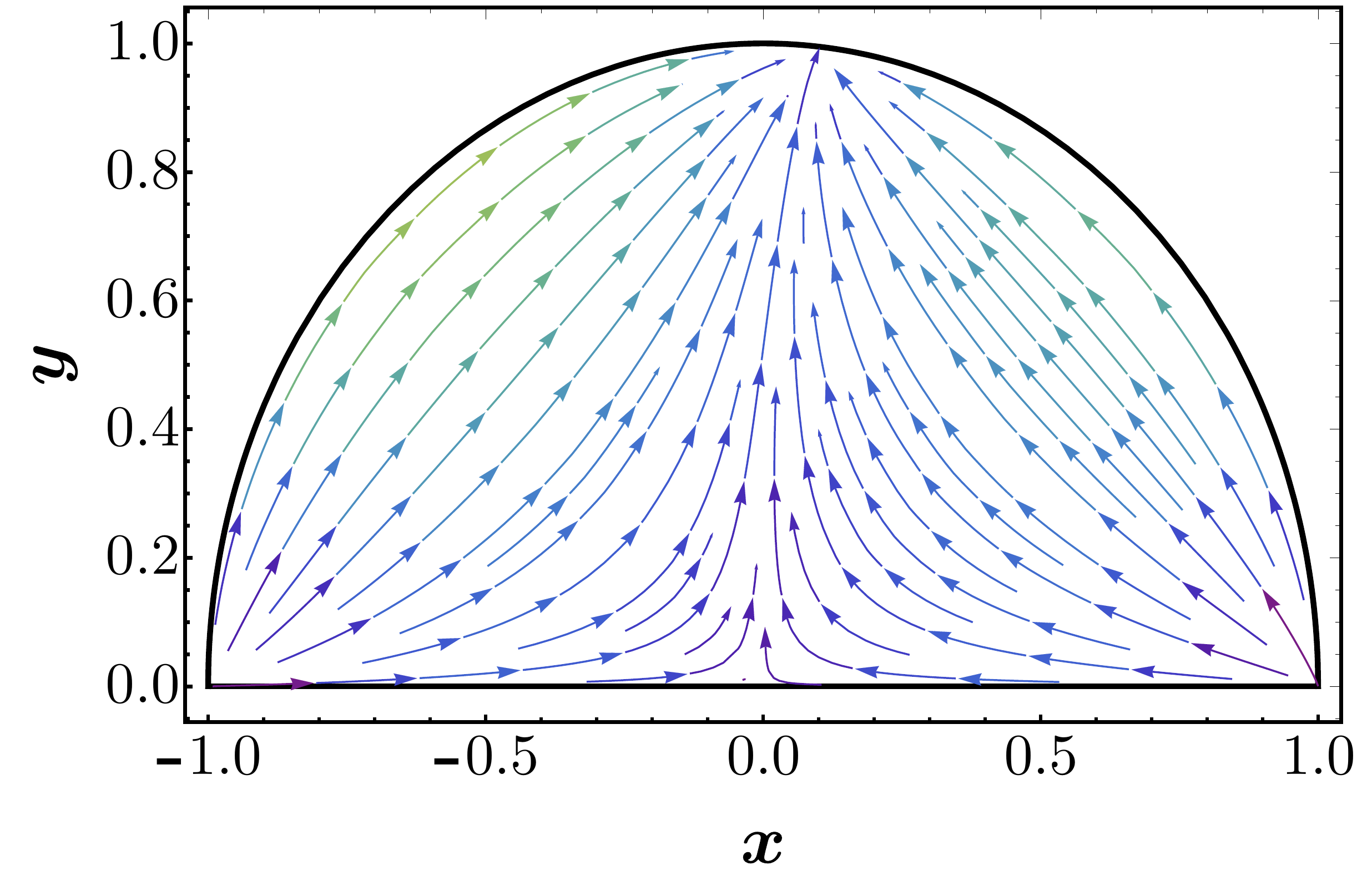}
\end{minipage}
\caption{Left panel: Evolution of various parameters in the future($N>0$); Right panel: $x-y$ phase space with a dark energy dominated attractor point for $C=0.6$.}
\label{fig:xypos}
\end{figure}

\begin{figure}[!htb]
\begin{minipage}[b]{.5\textwidth}
\includegraphics[scale=0.45]{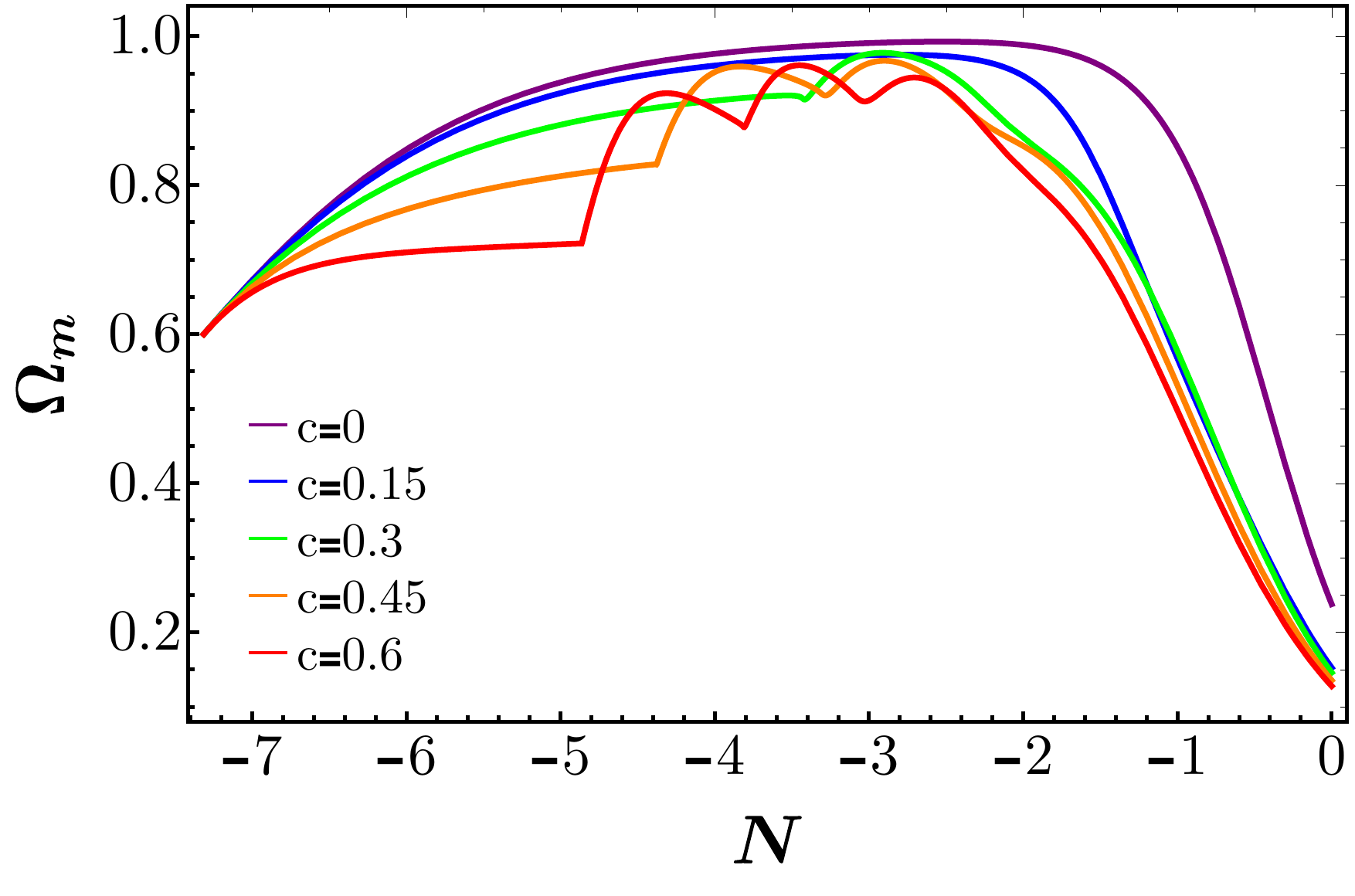}
%\caption{$k=5H_0$}\label{label-k5}
\end{minipage}\hfill
\begin{minipage}[b]{.5\textwidth}
\includegraphics[scale=0.45]{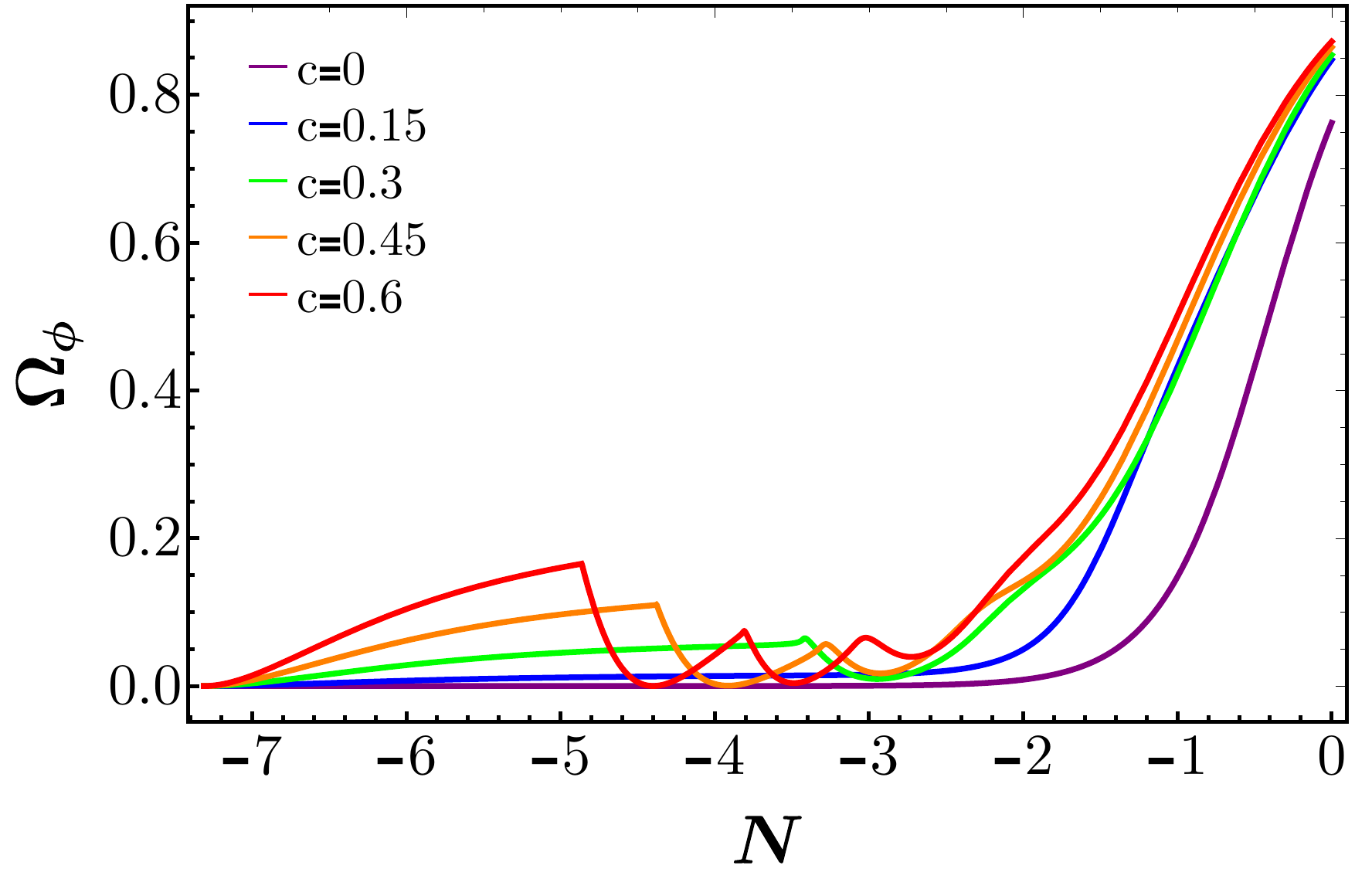}
\end{minipage}
\caption{Evolution of Energy density parameters as function of $N$; Left panel: Dark matter; Right panel: Dark energy.}
\label{fig:ommphi}
\end{figure}

In the previous subsection, we showed that the interacting dark sector model has a stable attractor solution, which corresponds to the late-time accelerated Universe. To demonstrate this point, we fix the variables other than $x$ and $y$ to be constants with the values of corresponding functions at $N = 7$. As we see in the left panel of Fig.~\eqref{fig:xypos}, this is a reasonable assumption since the relevant variables are nearly constant for $N>3$.  From the right panel of Fig.~\eqref{fig:xypos}, we see that a large range of $x$ and $y$ parameters lead to dark energy dominated attractor. It has to be noted that this is a rough representation of the phase space evolution since the other parameters in the system are slowly varying. For simplicity, we have assumed them to be constants while plotting the phase space diagram. Hence, the attractor in the phase space diagram is an instantaneous attractor. It is important to note that various potentials, including the one we have considered here, have been shown to have dark energy dominated attractor in the non-interacting scenario~\cite{2015-Amendola.Tsujikawa-DarkEnergyTheory,2001-Ng.etal-Phys.Rev.D}. Fig. \eqref{fig:ommphi} contains the plots of the evolution of energy density parameters for dark matter and dark energy. Figures point that different coupling strengths lead to a dark energy dominated Universe, and the dark energy dominated phase starts earlier as compared to the non-interacting scenario.
\begin{figure}[!htb]
\begin{minipage}[b]{.48\textwidth}
\includegraphics[scale=0.45]{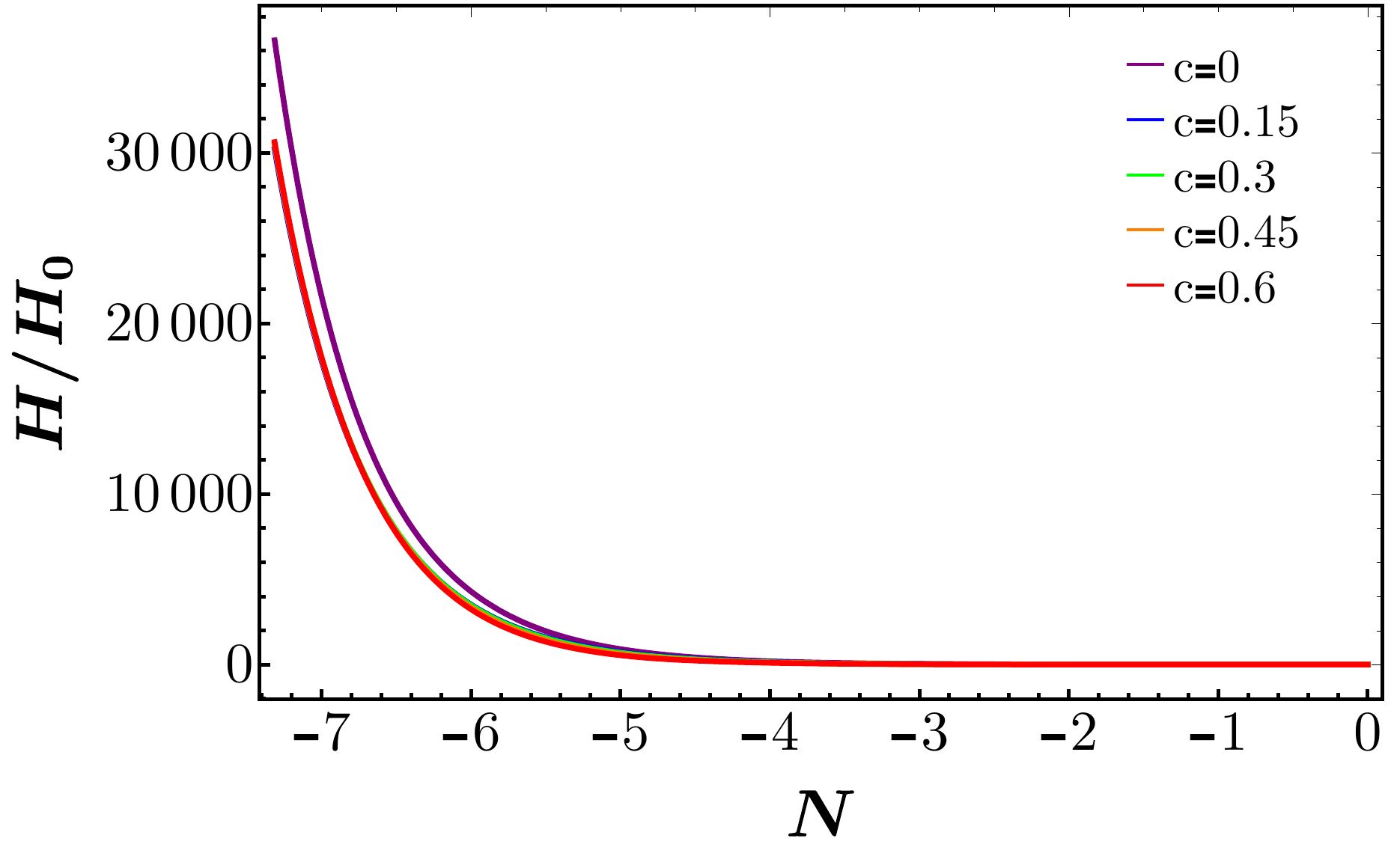}
\end{minipage}\hfill
\begin{minipage}[b]{.48\textwidth}
\includegraphics[scale=0.45]{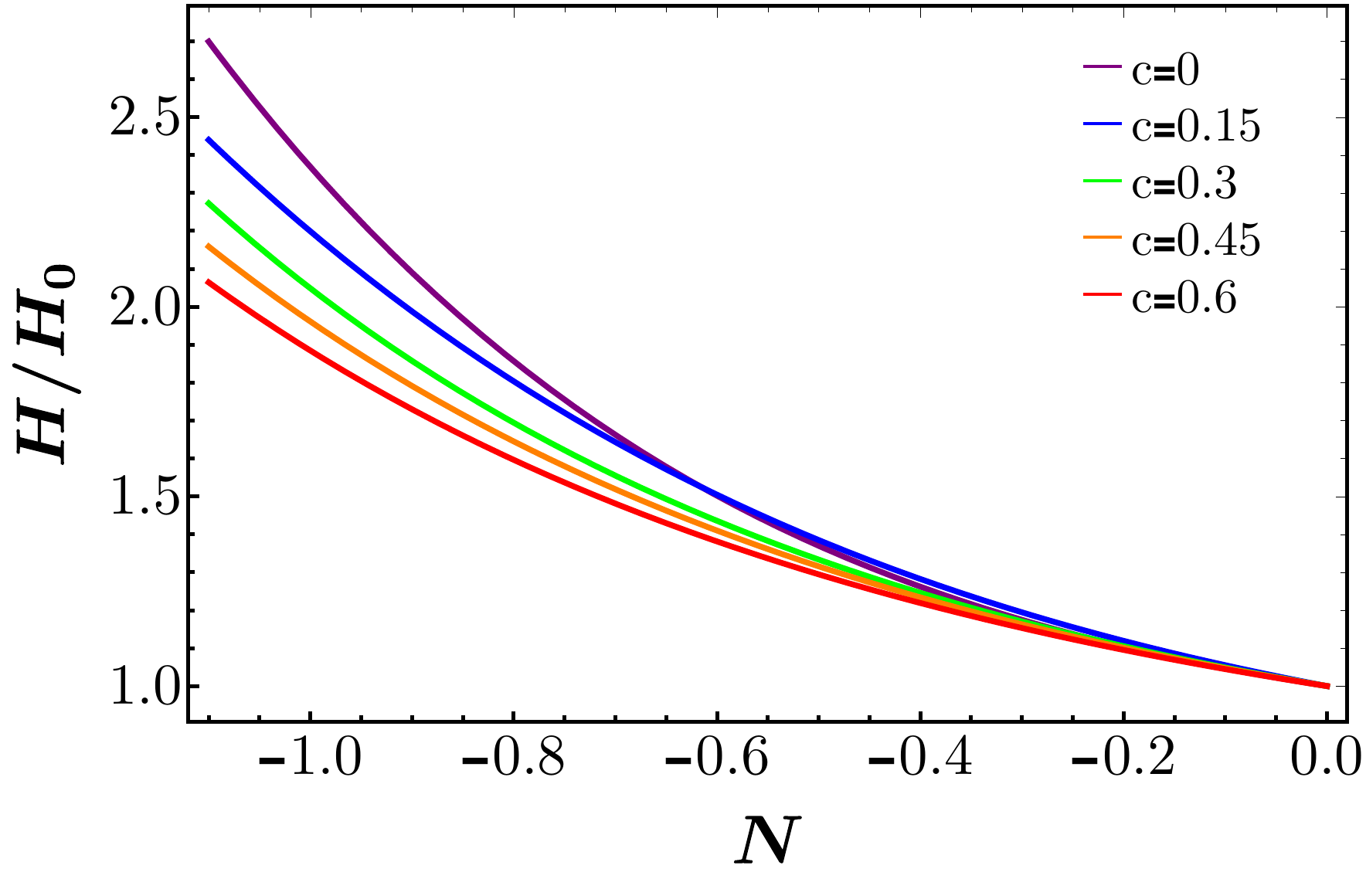}
\end{minipage}
\caption{Evolution of Hubble parameter $h=H/H_0$ as function of $N$.}
\label{fig:HbyH0}
\end{figure}
To further investigate, in Fig. \eqref{fig:HbyH0} we plot the evolution of scaled Hubble parameter $h=H/H_0$.
The right panel plot (for redshift range $2 < z  < 0$)  is the zoomed version of the left panel plot (for redshift range $1500 < z  < 0$). From the plots, it is clear that while the dark energy dominates early, the scaled Hubble parameter is less in the interacting dark-sector models than the non-interacting dark-sector model.  

\subsubsection{Scenario II : $C \leq 0$}

\begin{figure}[!htb]
\begin{minipage}[b]{.5\textwidth}
\includegraphics[scale=0.45]{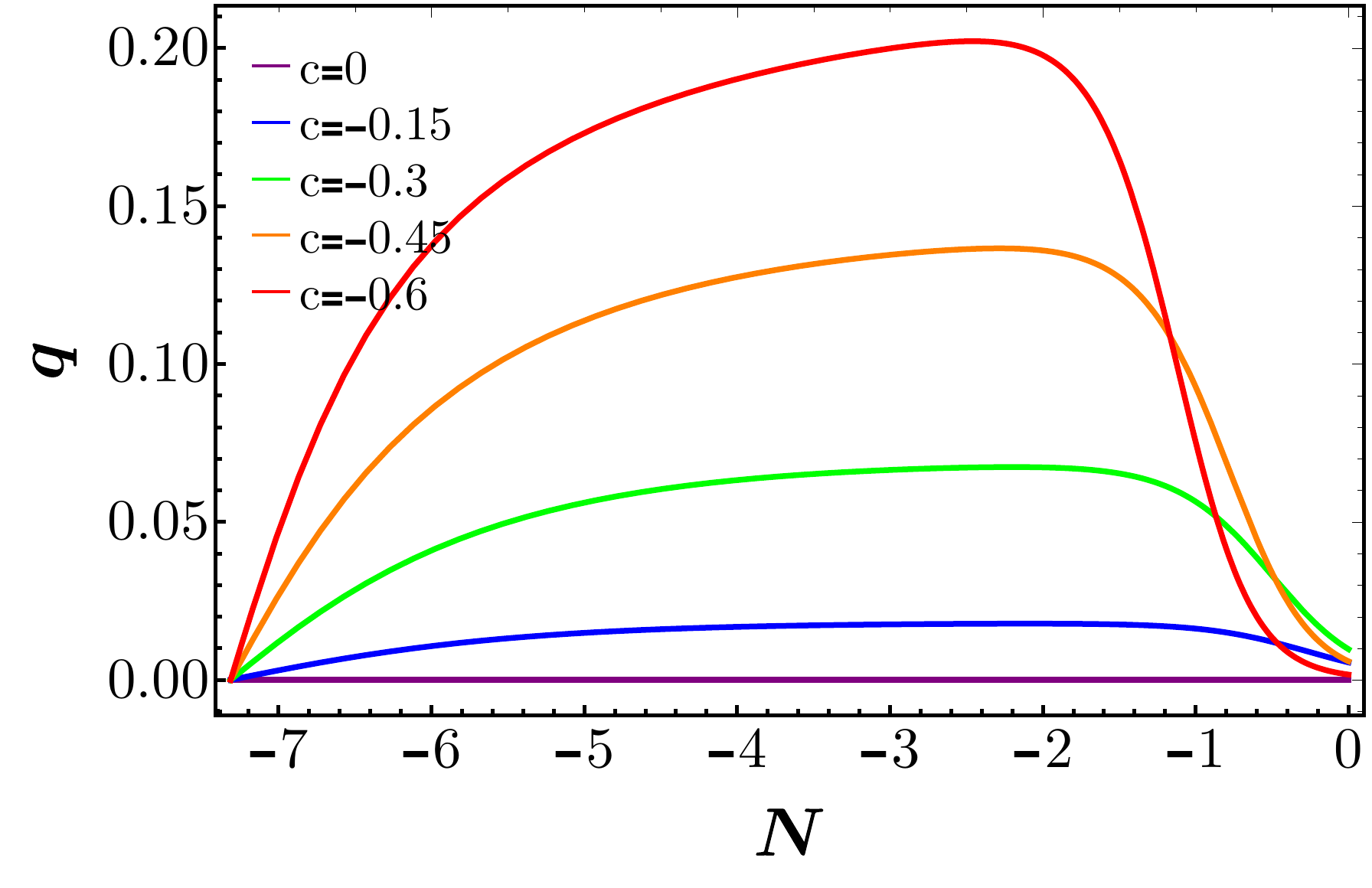}
%\caption{$k=5H_0$}\label{label-k5}
\end{minipage}\hfill
\begin{minipage}[b]{.5\textwidth}
\includegraphics[scale=0.45]{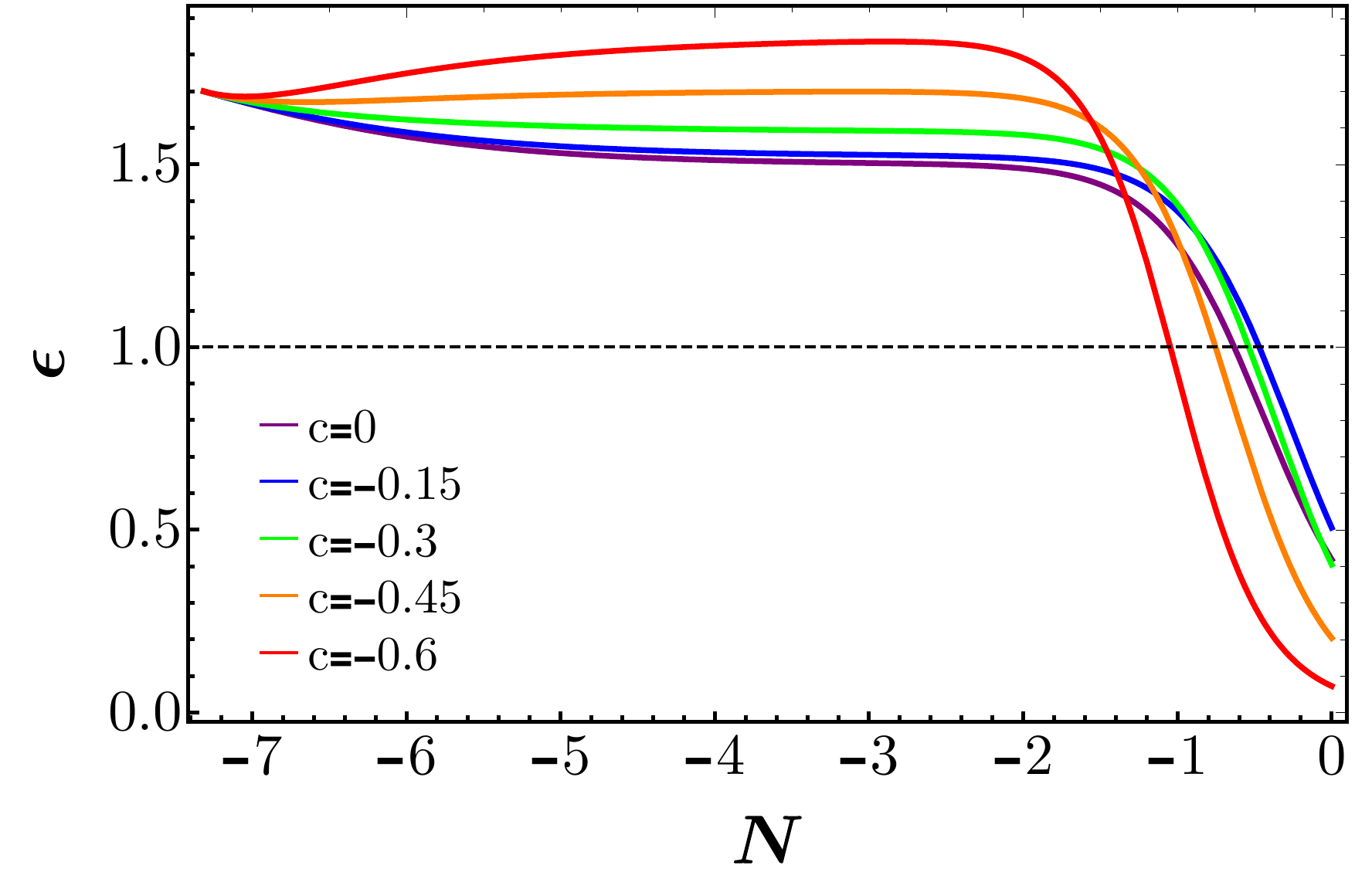}
\end{minipage}
\caption{Left panel: Evolution of interaction term $q \equiv \alpha \beta x \Omega_{m}$ as a function of $N$; Right panel: slow-roll parameter $\epsilon$ as function of $N$}
\label{fig:qepsilonII}
\end{figure}

In this scenario, the initial value of the coupling function is non-negative. Fig. \eqref{fig:qepsilonII} contains 
the plots of the evolution of the scaled interaction term $q$ (defined in Eq. \ref{eq:qQdef}) and slow-roll parameter $\epsilon$ (defined in Eq. \ref{def:epsilon}). As in the earlier scenario, the evolution with $C<0$ leads to accelerated expansion, and the interaction function stays positive during the evolution. However, the late-time evolution of these parameters is different from that of the $C \geq 0$ scenario.

\begin{figure}[!htb]
\begin{minipage}[b]{.5\textwidth}
\includegraphics[scale=0.32]{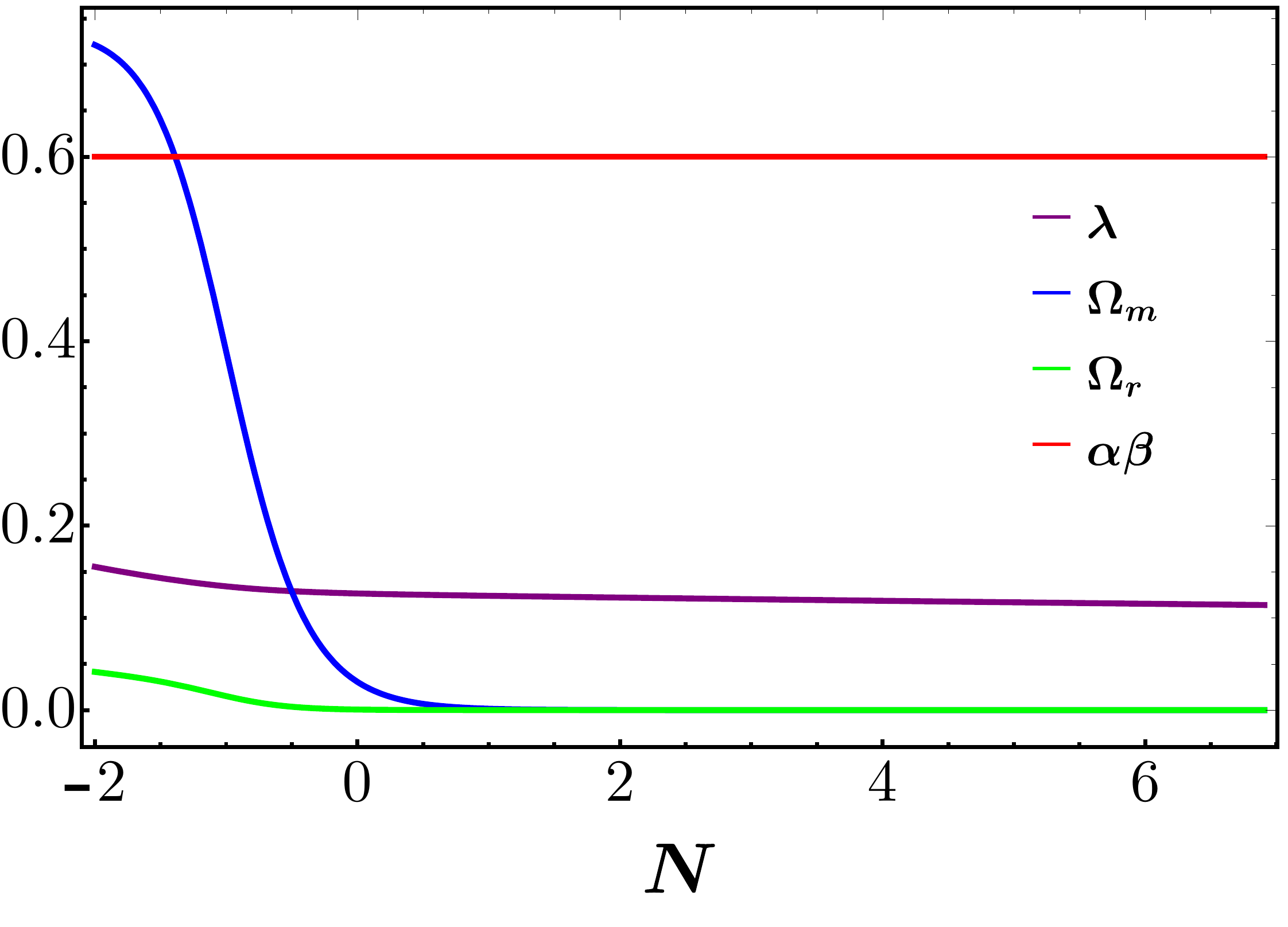}
%\caption{$k=5H_0$}\label{label-k5}
\end{minipage}\hfill
\begin{minipage}[b]{.5\textwidth}
\includegraphics[scale=0.32]{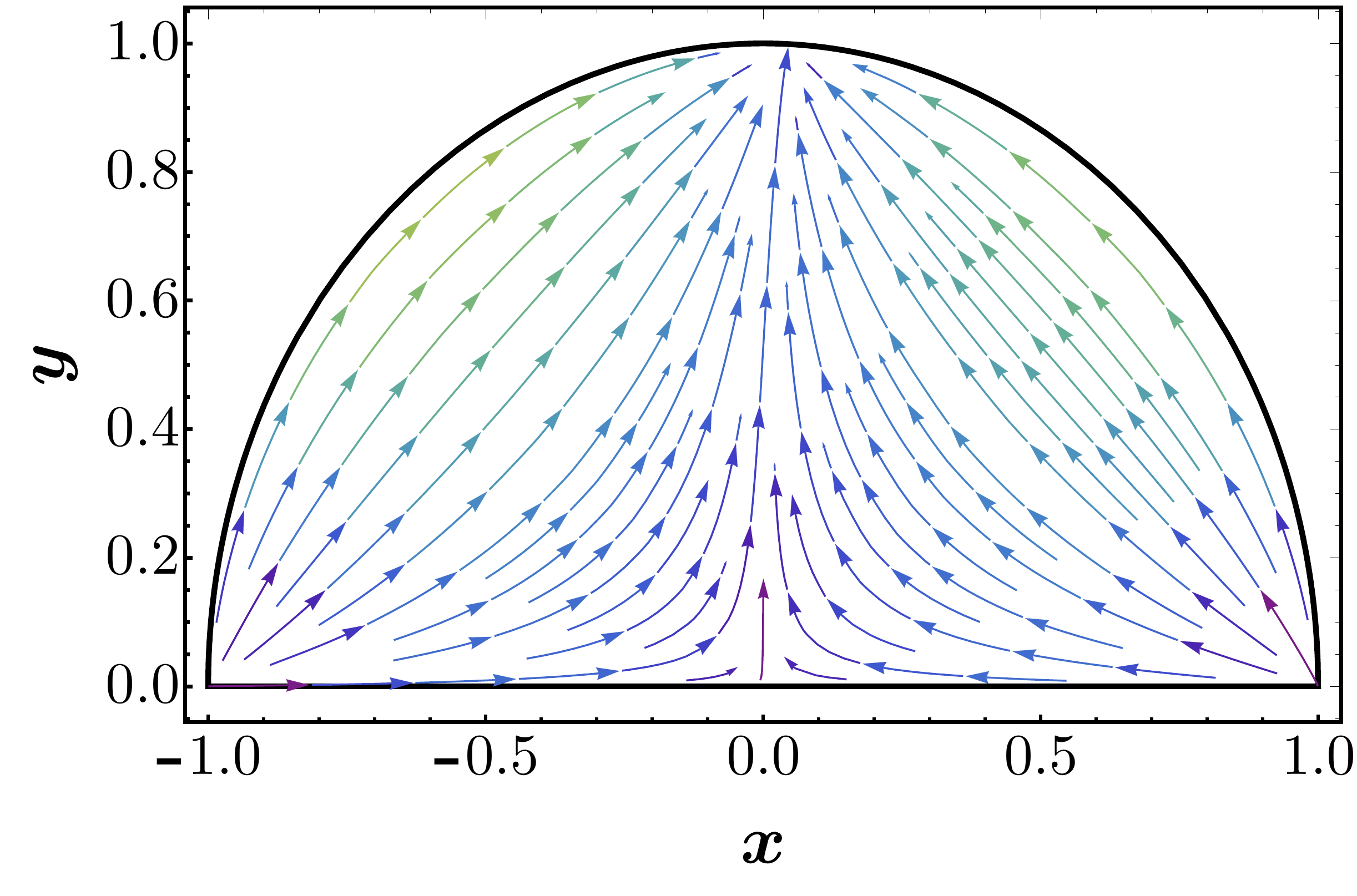}
\end{minipage}
\caption{ Left panel: Evolution of various parameters in the future($N>0$); Right panel: $x-y$ phase space with a dark energy dominated attractor point for $C=-0.6$.}
\label{fig:ommphineg}
\end{figure}

\begin{figure}[!htb]
\begin{minipage}[b]{.5\textwidth}
\includegraphics[scale=0.45]{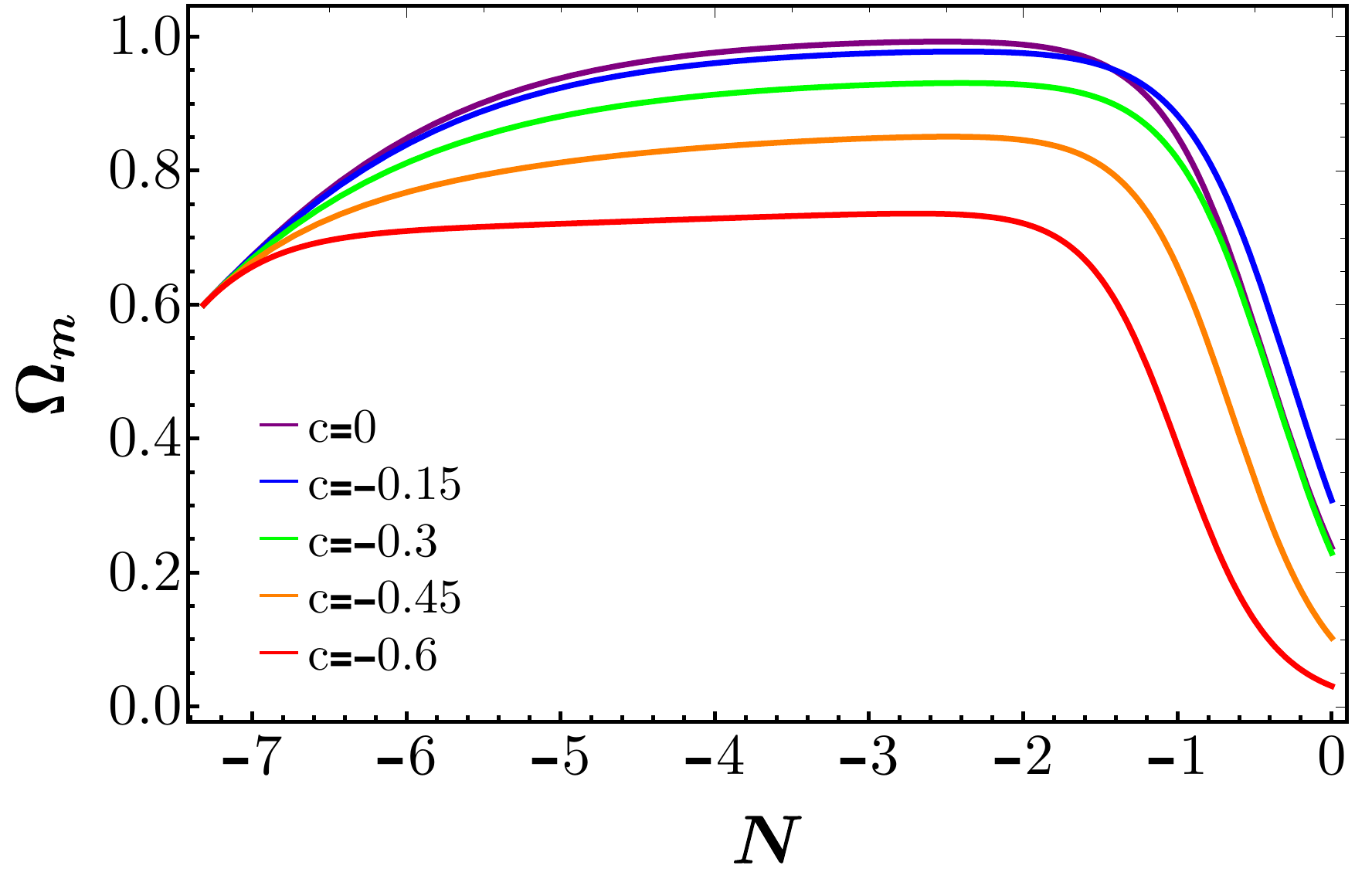}
%\caption{$k=5H_0$}\label{label-k5}
\end{minipage}\hfill
\begin{minipage}[b]{.5\textwidth}
\includegraphics[scale=0.45]{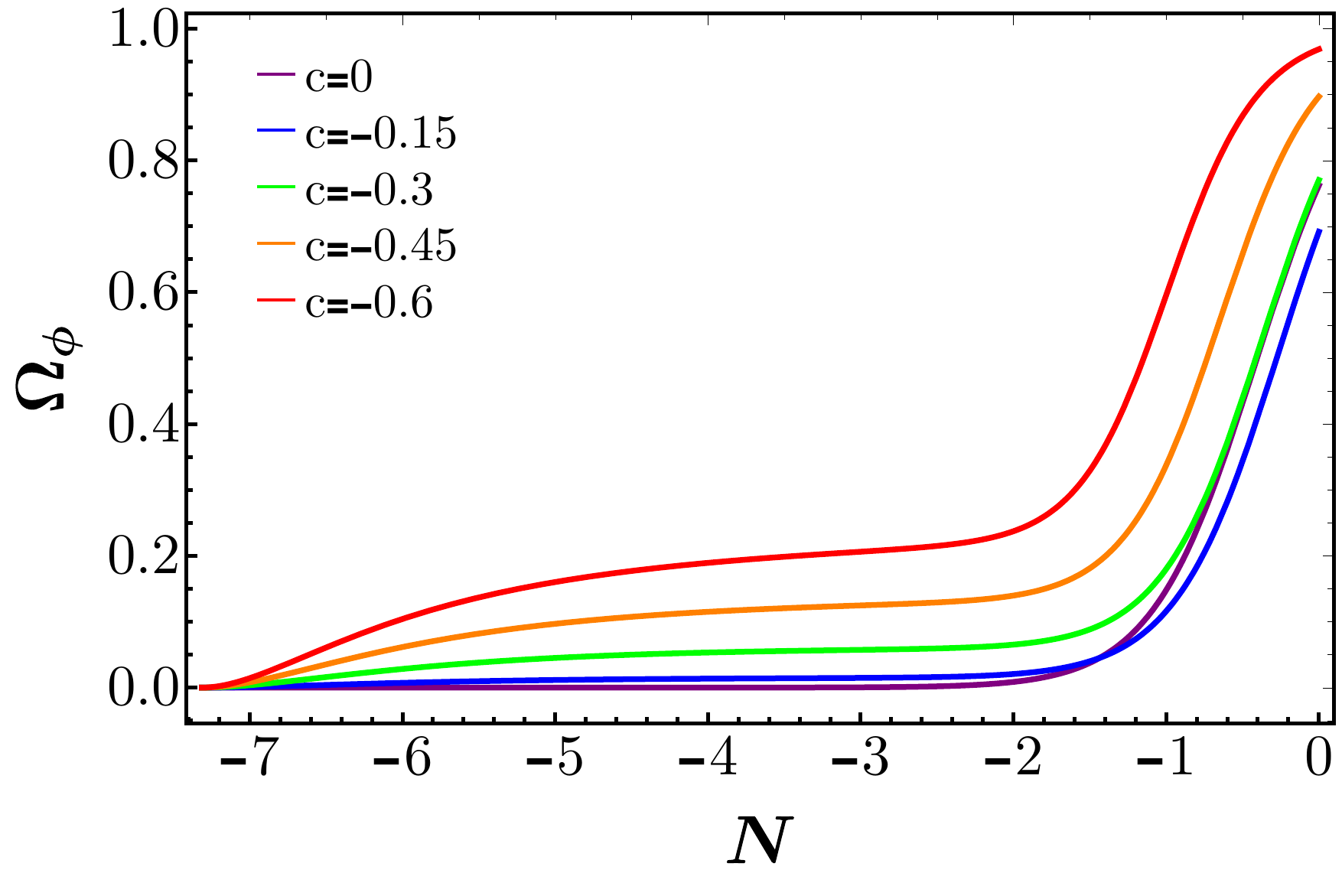}
\end{minipage}
\caption{Evolution of Energy density parameters as function of $N$; Left panel: Dark matter; Right panel: Dark energy.}
\label{fig:ommphiII}
\end{figure}

\begin{figure}[!htb]
\begin{minipage}[b]{.48\textwidth}
\includegraphics[scale=0.45]{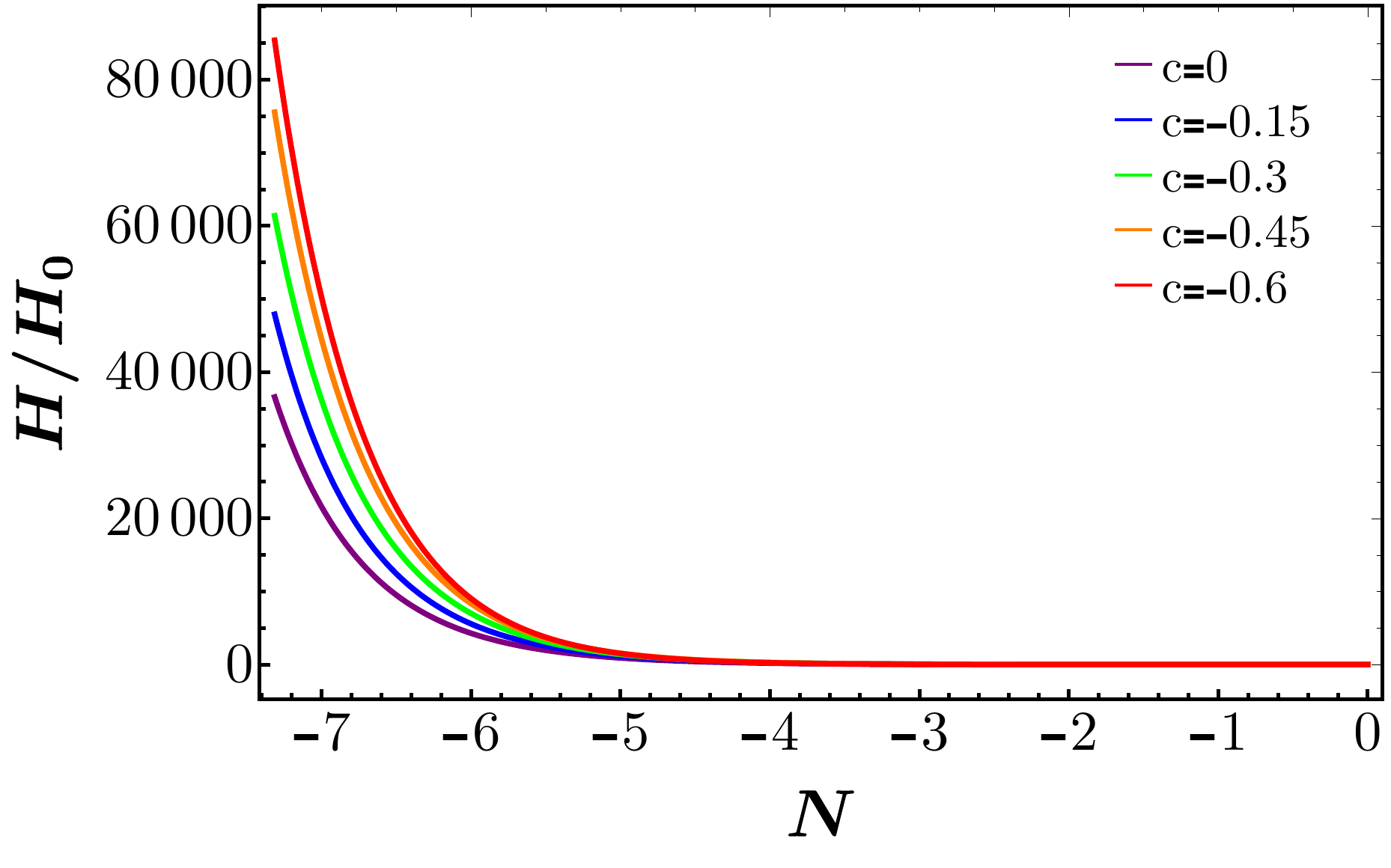}
\end{minipage}\hfill
\begin{minipage}[b]{.48\textwidth}
\includegraphics[scale=0.45]{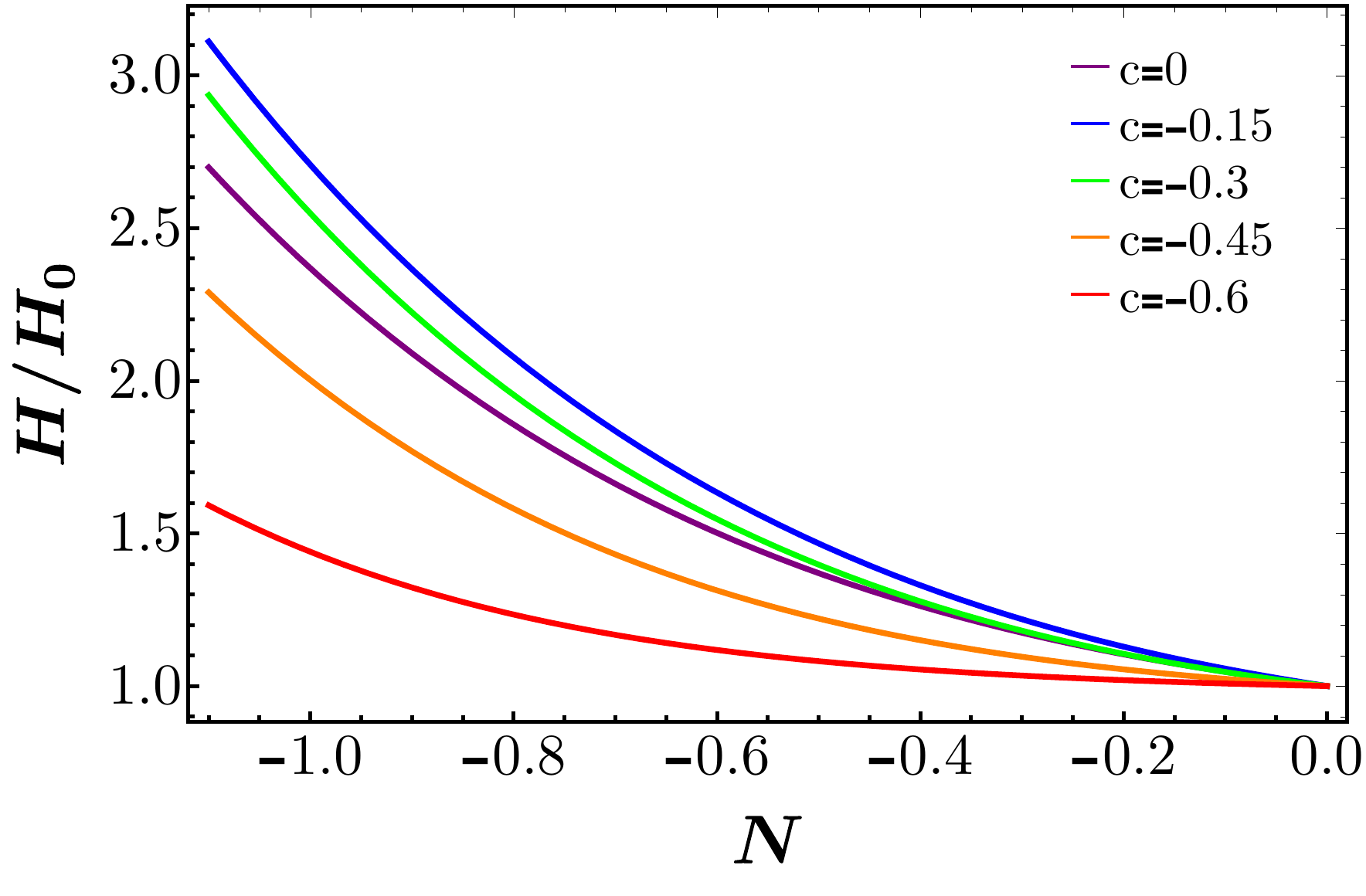}
\end{minipage}
\caption{Evolution of Hubble parameter $h=H/H_0$ as function of $N$.}
\label{fig:hubneg}
\end{figure}

Like in $C \geq 0$ scenario, Fig. \eqref{fig:ommphineg} contains the evolution of the system in the $x-y$ phase plane. This scenario also leads to stable dark energy dominated attractor. 
Fig. \eqref{fig:ommphiII} contains the evolution of $\Omega_{\phi}$ and $\Omega_m$. Like in the earlier scenario, the plots show the dark energy dominated phase in the late Universe. However, the late-time evolution of these parameters is different from the $C \geq 0$ scenario. Fig. \eqref{fig:hubneg}) also shows a similar trend in the evolution of the Hubble parameter. 

\section{Conclusions and Discussions}

In this work, we have constructed the dark energy - dark matter interaction from a classical field theory action. This action is obtained from the $f(\tilde{R},\tilde{\upchi})$ action using a conformal transformation and redefinition of the scalar fields. While the total energy-momentum tensor is conserved due to interaction, the energy-momentum of the individual components in the dark sector is not satisfied. This lead to an unique interaction term $Q^{\rm (F)}_{\nu}$. 

While the field theory description helps us to obtain the interaction from the action principle, the fluid description turns out to be more useful for analyzing cosmological observations. In that regard, the most common description of the interacting dark sector is in terms of dark matter fluid. However,  in the phenomenological description of the dark matter fluid interaction, there is no unique form of ${Q}_{\nu}$. In many of the models in the literature, the interaction strength $Q_{\nu}$ in the dark sector is introduced by hand. We have systematically shown that the one-to-one correspondence between the fluids and fields is possible only if the interaction term is given by $Q^{\rm (F)}_{\nu}$. In this specific case, the equations in the field theory description can be obtained from the fluid equations by a simple substitution of the variables. 

We classified the interacting dark energy models considered in the literature into two categories based on the field-theoretic description.  While many of the models have a field-theoretic description, many of the dark matter fluid interaction models do not have a field-theoretic description used in this work. The field-theoretic description used in this work is the simplest possible. It may be possible that by considering a generalized action, like Horndeski Lagrangian, some of these models may have a field-theoretic description~\cite{2020-Kase.Tsujikawa-}. This needs further investigation.

We defined a set of dimensionless variables and constructed a novel autonomous system that describes the evolution in a general quintessence dark energy interacting with dark matter. Studying the fixed points of the autonomous system, we showed that the interacting dark sector model has a stable attractor solution that describes the late-time accelerated Universe. As an example, we considered the model with $U(\phi) \propto 1/\phi$ and $\alpha(\phi) \propto \phi$. We have shown that a stable, dark energy dominated solution exists for this model for a range of coupling strengths (both positive and negative).

While the form of the interaction term ($Q^{\rm (F)}_{\nu}$) is unique, it still contains unknown functions like $\alpha(\phi)$, $\upchi$ and $V(\upchi)$. These can be constrained using particle physics models ~\cite{2016-DAmico.etal-Phys.Rev.D}.
The immediate question that arises is whether one can use some other tools to constrain further the suitable dark matter-dark energy model from the observations. This is currently under investigation.

\section{Acknowledgements}
We thank T. Padmanabhan for fruitful discussions and for drawing us into the dark sector dynamics! JPJ thanks M. Trodden for clarifications on Ref. \cite{2018-CarrilloGonzalez.Trodden-Phys.Rev.D}. Authors thank Archana Sangwan for the discussions. JPJ is supported by CSIR Senior Research Fellowship, India. The work is partially supported by IRCC Seed Grant, IIT Bombay and  ISRO-RESPOND grant.

\appendix
\section{Field theoretic formulation of the dark energy - dark matter interaction}
\label{app:A}
Consider the following Jordan frame action:
\begin{equation}
\label{eq:fRactiona}
 S_{J}=\int d^{4}x\sqrt{-\tilde{g}}\left[\frac{1}{2 \kappa^2} f(\tilde{R},\tilde{\upchi})- \frac{1}{2}   \tilde{g}^{\mu \nu} \tilde{\nabla}_{\mu} \tilde{\upchi} \tilde{\nabla}_{\nu} \tilde{\upchi} -V(\tilde{\upchi})\right] 
\end{equation}
where $f(\tilde{R},\tilde{\upchi})$ is arbitrary, smooth function of Ricci scalar 
($\tilde{R}$) defined in the 4-D metric $\tilde{g}_{\mu\nu}$, and 
the scalar field $\tilde{\upchi}$. Under conformal transformation and redefining the scalar fields, one can bring it to the Einstein frame with two interacting scalar fields~\cite{2001-Starobinsky.etal-Nucl.Phys.B}.

To keep calculations tractable, we assume the form of $f(\tilde{R},\tilde{\upchi})$ as 
\begin{equation}
 f(\tilde{R},\tilde{\upchi}) =   h(\tilde{\upchi})  f(\tilde{R})
\end{equation}
The above action can be rewritten as:
\begin{equation}
\label{eq:fRaction-02}
    S= \int d^{4}x \sqrt{-\tilde{g}} \left[ h(\tilde{\upchi})
    \left(  \frac{F \tilde{R}}{2 \kappa^2}-U\right) 
    - \frac{1}{2}\tilde{g}^{\mu \nu} \partial_{\mu}\tilde{\upchi} \partial_{\nu}\tilde{\upchi} - V(\tilde{\upchi})\right] 
\end{equation}
where 
\[
 F = \frac{\partial f}{\partial \tilde{R}} \quad \mbox{and} \quad 
 \tilde{U}=\frac{F \tilde{R}-f}{2 \kappa^2}
 \]
Under the following conformal transformation
\begin{equation}
    \hat{g}_{\mu \nu}= F \, \tilde{g}_{\mu \nu}
\end{equation}
the above action \eqref{eq:fRaction-02} becomes
\begin{eqnarray}
S &=& \int d^{4}x \sqrt{-\hat{g}} \left[ h(\tilde{\upchi})  \frac{\hat{R}}{2 \kappa^2}-h(\tilde{\upchi})\hat{U}+h(\tilde{\upchi})\sqrt{\frac{3}{2 \kappa^2}}\hat{\Box}\psi- \frac{h(\tilde{\upchi})}{2}\hat{g}^{\mu \nu} \partial_{\mu}\psi \partial_{\nu}\psi \right. \nonumber \\
%%%%
& & \qquad \qquad 
- \left. \frac{e^{-\sqrt{\frac{2 \kappa^2}{3}}\psi}}{}\hat{g}^{\mu \nu} \partial_{\mu}\tilde{\upchi} \partial_{\nu}\tilde{\upchi} - \hat{V}(\tilde{\upchi}) \right] 
\label{eq:conformaction}
\end{eqnarray}
where 
\begin{equation}
    \psi=\sqrt{\frac{3}{2 \kappa^2}} \ln \, F, \quad \hat{U}=\frac{\tilde{U}}{F^{2}}, \quad \hat{V}=\frac{V}{F^{2}} \, .
\end{equation}
Introducing one more conformal transformation:
\begin{equation}
    g_{\mu \nu}=h(\tilde{\upchi)} \hat{g}_{\mu \nu}
\end{equation}
the above action can be rewritten as:
\begin{eqnarray}
 S &=&  \int d^{4}x \sqrt{-g} \left[  \frac{R}{2\kappa^2}- \left[ \frac{1}{2he^{\sqrt{\frac{2 \kappa^2}{3}}\psi}}+\frac{3h_{,\tilde{\chi}}^{2}}{4 \kappa^2 h^{2}} \right] g^{\mu \nu} \partial_{\mu}\tilde{\upchi} \partial_{\nu}\tilde{\upchi}  \right. \nonumber \\
%%%%
& & \qquad \qquad  \left.  
- \frac{1}{2}g^{\mu \nu} \partial_{\mu}\psi \partial_{\nu}\psi-\sqrt{\frac{3}{2 \kappa^2}}\frac{h_{,\tilde{\chi}}}{h} g^{\mu \nu}\partial_{\mu}\tilde{\upchi}\partial_{\nu}\psi - \hat{W}\right]  \, ,
    \label{eq:transaction}
\end{eqnarray}    
where
\begin{equation}
\label{eq:transpotential}
    \hat{W}=\frac{FR-f}{\kappa hF^{2}}+\frac{V}{h^{2}F^{2}}
\end{equation}
The above action in the Einstein frame neatly separates into 
Ricci scalar and the scalar fields. However, the scalar fields are not 
in canonical form. Since the metric $g_{\mu \nu}$ appears in all the 
kinetic part of the scalar fields, the field space line-element can be
written as: 
\begin{equation}
 d\ell^2=  \left[ \frac{1}{he^{\sqrt{\frac{2 \kappa^2}{3}}\psi}}+\frac{3h_{,\tilde{\chi}}^{2}}{ 2 \kappa^2 h^{2}} \right] d\tilde{\upchi}^{2}+2\sqrt{\frac{3}{2 \kappa^2}}\frac{h_{,\tilde{\chi}}}{h} d\tilde{\upchi} d\psi + d\psi^2
 \label{eq:fieldspace01}
\end{equation}
It has to be noted that it is impossible to bring the above line element to Euclidean form by the redefinition of the fields. Thus, the field-space line-element can be written in many different ways, leading to a different interaction between the two scalar fields. We list two cases below:
\begin{enumerate}
\item One of the simplest options is to redefine the fields as~\cite{2019-Johnson.etal-Gen.Rel.Grav.}:
\begin{equation}
    \sqrt{\frac{3}{2 \kappa^2}} \ln h+\psi = \phi, \quad \tilde{\upchi} = \upchi
\end{equation}
Then the field space line element \eqref{eq:fieldspace01} reduces to:
\begin{equation}
    d\ell^{2 }=\frac{1}{h e^{\sqrt{\frac{2 \kappa^2}{3}}\psi}}d\tilde{\upchi}^{2}+\left[ d(\sqrt{\frac{3}{2 \kappa^2}}lnh+\psi)\right] ^{2} = e^{-\sqrt{\frac{2 \kappa^2}{3}}\phi} d \upchi^2 + d\phi^2
    \label{eq:fieldspace02} 
\end{equation}
Under this field redefinition, the Einstein frame action \eqref{eq:transaction} is given by:
\begin{equation}
\label{eq:fRconf}
 S_{E} =\int d^{4}x\sqrt{-g}\left[\dfrac{R}{2\kappa^2}-\dfrac{1}{2}g^{\mu \nu}\nabla_{\mu}\phi \nabla_{\nu}\phi-U(\phi)-\dfrac{1}{2}e^{-\sqrt{\frac{2\kappa^2}{3}}\phi}g^{\mu \nu}\nabla_{\mu}\upchi\nabla_{\nu}\upchi-e^{-2\sqrt{\frac{2\kappa^2}{3}}\phi}V(\upchi) \right],
\end{equation} 

\item Let us now consider the following redefinition of the fields: 
\begin{eqnarray}
\nonumber
e^{2\alpha(\phi)} \left(\dfrac{\partial \upchi}{\partial \psi}\right)^2 + \left(\dfrac{\partial \phi}{\partial \psi}\right)^2  & = & 1
\\
\nonumber
e^{2\alpha(\phi)} \dfrac{\partial \upchi}{\partial \tilde{\upchi}} \dfrac{\partial \upchi}{\partial \psi} + \dfrac{\partial \phi}{\partial \tilde{\upchi}} \dfrac{\partial \phi}{\partial \psi} & = & \sqrt{\dfrac{3}{2 \kappa^2}} \dfrac{h_{,\tilde{\
chi}}}{h}
\\
e^{2\alpha(\phi)}\left(\dfrac{\partial \upchi}{\partial \tilde{\upchi}}\right)^2 + \left(\dfrac{\partial \phi}{\partial \tilde{\upchi}}\right)^2  & = & \dfrac{1}{h e^{\sqrt{\frac{2 \kappa^2}{3}}\psi}} + \dfrac{3}{2 \kappa^2} \dfrac{h_{,\tilde{\chi}}^2}{h^2}
\end{eqnarray}
where $\upchi \equiv \upchi(\tilde{\upchi},\psi)$, $\phi \equiv \phi(\tilde{\upchi},\psi)$, and $\alpha(\phi)$ is an arbitrary 
function of $\phi$. Under this redefinition, the field space line-element  
\eqref{eq:fieldspace01} reduces to: 
\begin{equation}
ds^2 = e^{2\alpha(\phi)} d \upchi^2 + d \phi^2
\end{equation}
Thus, the Einstein frame action takes the form. 
\begin{equation}
S_{E} = \int d^4x \sqrt{-g}\left(\dfrac{1}{2 \kappa^2}R-\dfrac{1}{2}g^{\mu \nu}\nabla_{\mu}\phi \nabla_{\nu}\phi - U(\phi)-\dfrac{1}{2} e^{2\alpha(\phi)}g^{\mu \nu}\nabla_{\mu}\upchi \nabla_{\nu}\upchi -e^{4 \alpha(\phi)} V(\upchi) \right).
\end{equation}
and is identical to the action \eqref{eq:Scde} in Sec. \eqref{sec:2}. This 
action describes two interacting scalar fields with an arbitrary coupling represented by the function $\alpha(\phi)$. 
\end{enumerate}
\section{Cosmological evolution in Newtonian gauge}
\label{App2}

Consider the spatially flat FRW metric with first order scalar perturbations in Newtonian gauge~\cite{2008-Weinberg-Cosmology}:
\begin{equation}
\label{eq:newtmetric}
g_{00}=-\left(1+2\Phi \right),\quad g_{0i}=0, \quad g_{ij} = a^2 (1-2\Psi)\delta_{ij}.
\end{equation}
where $a\equiv a(t)$ is the scale factor with Hubble parameter given by $H = \dot{a}/a$ and $\Phi\equiv \Phi(t,x,y,z)$ and $\Psi \equiv \Psi(t,x,y,z)$ are scalar perturbations. 

The scalar fields $\phi$ and $\upchi$, dark matter fluid energy density ($\rho_{m}$), dark matter fluid pressure ($p_m$) and the interaction strength ($Q_{\nu}$) can be split into background and perturbed parts as:
\begin{eqnarray}
\label{eq:rhompmdefn}
& & \phi = \overline{\phi}+\delta\phi ,\quad \upchi = \overline{\upchi} + \delta \upchi, \quad \rho_{m}
=\overline{\rho}_m+\delta \rho_m, \quad p_m = \overline{p}_m+\delta p_m, 
\quad Q_{\nu} = \overline{Q}_{\nu} + \delta Q_{\nu} \\
\label{eq:dmfourvn}
& & u_{\mu} = \overline{u}_{\mu}+\delta u_{\mu}, \quad \overline{u}_0 = -1, \quad \delta u_0 = -\Phi , \quad \overline{u}_i=0 , \quad \delta u_i = \dfrac{\partial \delta u^s}{\partial x^i} , \quad \delta u^s = -\dfrac{\delta \upchi}{\dot{\overline{\upchi}}}
\end{eqnarray}
In the following subsection, we present the evolution equations for the first-order perturbations in Newtonian gauge.

\subsection{Correspondence between fields and fluids in first order perturbations}
\label{App:2b}

In the fluid description, the first order scalar perturbations, in Newtonian gauge, satisfy the following equations~\cite{2008-Weinberg-Cosmology}:
\begin{eqnarray}
 \Psi -\Phi & = & 0  \\
\dot{\Psi} + H \Phi  &=& \dfrac{\kappa^2}{2} \left[ \dot{\bar{\phi}}\delta \phi - \left(\bar{\rho}_m + \bar{p}_m \right) \delta u^s \right] \\
3 H \dot{\Psi} - \dfrac{\nabla^2 \Psi}{a^2} + 3 H^2 \Phi &=& -\dfrac{\kappa^2}{2} \left( \delta \rho_m + \dot{\delta\phi} \dot{\bar{\phi}} - \Phi \dot{\bar{\phi}}^2 + U_{,\phi}(\overline{\phi})\delta\phi \right) \\
\nonumber
3\ddot{\Psi} + \dfrac{\nabla^2 \Phi}{a^2} + 6 \Phi \left( H^2 + \dot{H} \right) + 3 H \left( 2\dot{\Psi} + \dot{\Phi} \right) &=& \dfrac{\kappa^2}{2} \left( \delta\rho_m + 3 \delta p_m + 4\dot{\delta \phi} \dot{\bar{\phi}} - 4 \Phi \dot{\bar{\phi}}^2 - 2 U_{,\phi}(\bar{\phi})\delta \phi \right)   \\
\end{eqnarray}
From Eq. \eqref{eq:IntDEDM}, the conservation equations for the dark energy field and dark matter fluid in the first order perturbations are given by:
\begin{eqnarray}
\nonumber
&  & \dot{\delta \rho_m} + 3 H (\delta p_m + \delta \rho_m)+(\overline{p}_m+\overline{\rho}_m)\left[\dfrac{\nabla^2 \delta u^s}{a^2} - 3 \dot{\Psi}\right] = -\delta Q \\
& & 
\dot{\overline{\phi}}\left(\ddot{\delta \phi}-\dfrac{\nabla^2 \delta \phi}{a^2}-2 \Phi \ddot{\overline{\phi}} + U_{,\phi \phi}(\overline{\phi}) \delta \phi \right) + \dot{\delta \phi} \left( \ddot{\overline{\phi}}+6 H \dot{\overline{\phi}} + U_{,\phi}(\overline{\phi}) \right) - \dfrac{\dot{\overline{\phi}}^2}{2} \left(3 \dot{\Psi} +\dot{\Phi} + 6 H \Phi  \right)  = \delta Q  .
\\
\end{eqnarray}
The above equations are generic equations for the coupled dark matter fluid
and dark energy field with arbitrary interaction term $\delta {Q}$. As discussed in Sec. \eqref{sec:2a}, starting from the Jordan frame action \eqref{eq:fRaction}, the interaction term $Q_{\nu}^{\rm (F)} $ in Eq. \eqref{eq:traceinter}  is uniquely written in terms of dark energy scalar field and dark matter fluid. In this case, the perturbed interaction term is given by
\begin{equation}
 \delta Q^{\rm (F)} =-(\delta \rho_m - 3 \delta p_m)\alpha_{,\phi}(\overline{\phi})\dot{\overline{\phi}}-(\overline{\rho}_m-3\overline{p}_m)\left[\alpha_{,\phi \phi}(\overline{\phi}) \dot{\overline{\phi}} \delta \phi + \alpha_{,\phi}(\overline{\phi}) \dot{\delta \phi}\right] 
\end{equation}
We now show that the above equations are consistent with the field theory description \emph{only} for this form of interaction ${Q}^{\rm (F)}$.  Substituting $\rho_{m}, p_{m},\delta \rho_{m}$, and $\delta p_{m}$ from Eq.~\eqref{eq:dmrhop}, the perturbed equations of motion for $\phi$ and $\upchi$, respectively, are:
\begin{eqnarray} 
\nonumber
& & \ddot{\delta \upchi}-\dfrac{\nabla^2 \delta \upchi}{a^2} + e^{2\alpha}V_{,\chi \chi}(\overline{\upchi}) \delta \upchi - \dot{\overline{\upchi}} \left( 3 \dot{\Psi} + \dot{\Phi} \right) + 2e^{2\alpha}V_{,\chi}(\overline{\upchi}) \Phi + 3 H \dot{\delta \upchi}
\\
& & \qquad \qquad + 2 \alpha_{,\phi}(\overline{\phi})
 \left(\dot{\overline{\phi}} \dot{\delta \upchi}
 + \dot{\overline{\upchi}} \dot{\delta \phi}\right) 
%%%%
  + 2 \delta \phi \left[ \dot{\overline{\phi}}\dot{\overline{\upchi}} \alpha_{,\phi \phi}(\overline{\phi}) + e^{2 \alpha} \alpha_{,\phi}(\overline{\phi}) V_{,\chi}(\overline{\upchi}) \right] =  0 
\\
%%%%%%%%%%%%%
\nonumber
& & \ddot{\delta \phi}-\dfrac{\nabla^2 \delta \phi}{a^2}+3H \dot{\delta \phi}  + U_{,\phi \phi}(\overline{\phi}) \delta \phi -  \dot{\overline{\phi}} \left( 3 \dot{\Psi} + \dot{\Phi} \right) + 2 U_{,\phi}(\overline{\phi}) \Phi + 2e^{2\alpha} \alpha_{,\phi}(\overline{\phi})\left[2 e^{2\alpha} V_{,\chi}(\overline{\upchi}) \delta \upchi - \dot{\overline{\upchi}} \dot{\delta \upchi} \right]  
\\
%%%
& &  \qquad \qquad + 2 e^{2\alpha} \alpha_{,\phi}(\overline{\phi})^2 \delta \phi \left[8 e^{2\alpha} V(\overline{\upchi}) - \dot{\overline{\upchi}}^2 \right]+e^{2\alpha}\alpha_{,\phi \phi}(\overline{\phi}) \delta \phi \left[4 e^{2\alpha}V(\overline{\upchi}) - \dot{\overline{\upchi}}^2 \right] = 0
\end{eqnarray}
The above perturbed field equations are identical to the equations obtained from Eqs. (\ref{eq:eomchi}, \ref{eq:eomphi}), respectively. The perturbed interaction term in the field theory picture also can be obtained by the direct substitution of the variables:
\begin{eqnarray}
\nonumber
\delta Q^{\rm (F)} &=& 2 e^{2\alpha}\alpha_{,\phi}(\overline{\phi}) \dot{\overline{\phi}}\left[\dot{\overline{\upchi}}\dot{\delta \upchi}-2 e^{2 \alpha} V_{,\chi}(\overline{\upchi}) \delta \upchi -\dot{\overline{\upchi}}^2\Phi \right] + e^{2\alpha} \alpha_{,\phi \phi}(\overline{\phi}) \dot{\overline{\phi}} \delta \phi \left[ \dot{\overline{\upchi}}^2 - 4 e^{2 \alpha} V(\overline{\upchi}) \right] 
\nonumber \\
&+& 2 e^{2 \alpha} \alpha_{,\phi}(\overline{\phi})^2 \dot{\overline{\phi}} \delta \phi \left[ \dot{\overline{\upchi}}^2 - 8 e^{2\alpha} V(\overline{\upchi})\right] + e^{2\alpha} \alpha_{,\phi}(\overline{\phi}) \dot{\delta \phi}\left[\dot{\overline{\upchi}}^2 - 4 e^{2 \alpha} V(\overline{\upchi}) \right]
\end{eqnarray}
\section{Background evolution in a general interacting dark energy-dark matter model}
\label{App:3}
In various quintessence models considered in the literature, the scalar field's dimensions differ depending on the nature of the potential, especially in the case of the power-law potentials~\cite{2013-Pavlov.etal-Phys.Rev.D}. To include those scenarios, we rewrite the action for the interacting dark sector 
\begin{equation}
\label{eq:genaction}
S = \int d^4x \sqrt{-g} \left[\dfrac{M_{Pl}^2}{2}R - C_1\left(\dfrac{1}{2}\nabla^{\mu}\phi\nabla_{\mu}\phi
+ U(\phi) \right) - C_2 \left( \dfrac{1}{2}e^{2\alpha(\phi)}\nabla^{\mu}\chi\nabla_{\mu}\chi + e^{4\alpha(\phi)}
V(\chi) \right) \right],
\end{equation}
where $C_1$ and $C_2$ are constants.
Then background energy density and pressure of the dark matter is defined by
\begin{equation}
\label{eq:rhomdef}
\rho_{m} = C_2 e^{2\alpha(\phi)} \left( \dfrac{\dot{\chi}^2}{2} + e^{2\alpha(\phi)}V(\chi) \right)
\end{equation}
\begin{equation}
\label{eq:pmdef}
p_m = C_2 e^{2\alpha(\phi)} \left( \dfrac{\dot{\chi}^2}{2} - e^{2\alpha(\phi)}V(\chi) \right)
\end{equation}
Then the energy conservation equations become
\begin{equation}
\label{eq:phieq}
C_1\left(\ddot{\phi} + 3H\dot{\phi} + U_{\phi}(\phi) \right)\dot{\phi} = Q
\end{equation}
\begin{equation}
\label{eq:rhomeq}
\dot{\rho_m} + 3H(\rho_m + p_m) = -Q,
\end{equation}
where the interaction term Q is given by
\begin{equation}
\label{eq:qeq}
Q = C_2 \alpha_{\phi}(\phi) \dot{\phi} \left( e^{2\alpha(\phi)} \dot{\chi}^2 - 4 e^{\alpha(\phi)}V(\chi)\right) = -\alpha_{\phi}(\phi) \dot{\phi}(\rho_m - 3p_m)
\end{equation}
Friedmann equations are given by
\begin{equation}
\label{eq:fried1}
H^2 = \dfrac{1}{3 M_{Pl}^2}\left[ \rho_m +C_1\left( \dfrac{\dot{\phi}^2}{2} + U(\phi)\right) \right]
\end{equation}
\begin{equation}
\label{eq:fried2}
2 \dot{H} = -\dfrac{1}{M_{Pl}^2}\left( \rho_m + p_m + C_1 \dot{\phi}^2 \right)
\end{equation}

%\bibliography{bibdata.bib}

\end{document}